\documentclass[aps,pra,superscriptaddress,twocolumn,floatfix,longbibliography]{revtex4-1}
\usepackage{dcolumn}
\usepackage{graphicx,color}
\usepackage[utf8]{inputenc}
\usepackage{amsmath,amssymb,bm,amsfonts,dsfont,mathrsfs,amsthm}
\usepackage{braket}
\usepackage{multirow}
\usepackage{txfonts,comment}
\usepackage[colorlinks=true,linkcolor=blue,citecolor=blue,urlcolor=blue]{hyperref}
\usepackage{soul}

\begin{document}

\title{Squeezed Phonon Lasing via Floquet-Controlled Solid-State Defects} 

\author{Hugo Molinares}
\affiliation{Departamento de Ciencias F\'{\i}sicas, Universidad de La Frontera, Casilla 54-D, Temuco, Chile}

\author{Gianluca Rastelli}
\affiliation{
Pitaevskii BEC Center, CNR-INO and Dipartimento di Fisica, Universita di Trento, I-38123 Trento, Italy}
\affiliation{
INFN-TIFPA, Trento Institute for Fundamental Physics and Applications, Via Sommarive 14, I-38123 Trento, Italy}

\author{Victor Montenegro}
\affiliation{Department of Applied Mathematics and Sciences, College of Computing and Mathematical Sciences, Khalifa University of Science and Technology, 127788 Abu Dhabi, United Arab Emirates}
\affiliation{Institute of Fundamental and Frontier Sciences, University of Electronic Science and Technology of China, Chengdu 611731, China}
\affiliation{Key Laboratory of Quantum Physics and Photonic Quantum Information, Ministry of Education, University of Electronic Science and Technology of China, Chengdu 611731, China}

\author{Vitalie Eremeev}  
\email{vitalie.eremeev@gmail.com}
\affiliation{Centro Multidisciplinario de F\'{\i}sica, Vicerrector\'{\i}a de Investigación, Universidad Mayor, 8580745 Santiago, Chile} 

\begin{abstract}
We propose a general Floquet-engineered scheme for squeezed phonon lasing that enables a continuous crossover from conventional phase-diffused phonon lasing to phase-locked squeezed phonon lasing. Our system consists of two pairs of periodically driven spins---a primary pair and an ancilla pair---coupled to a single mechanical mode. While the underlying mechanism applies generally to such spin-mechanical systems, we focus on a solid-state implementation based on color centers embedded in a circular hexagonal boron nitride (hBN) membrane. We show that the primary spins provide mechanical gain, whereas the ancilla pair engineer a dissipative channel for the Bogoliubov mode, thereby stabilizing robust steady-state squeezed phonon lasing. We support these findings through a comprehensive steady-state analysis of the lasing threshold, mechanical occupation, emission spectrum, and second-order correlations. Remarkably, we demonstrate that Floquet engineering intrinsically induces phase locking while enabling controllable quadrature squeezing, providing a simple and versatile route toward squeezed lasing in solid-state spin-mechanical platforms.
%
%
\end{abstract}

\maketitle

\section{Introduction}
\label{sec1}
The rich physics of phonon lasing---the acoustic analogue to optical laser emission---has been the subject of intensive theoretical research~\cite{ bargatin2003nanomechanical, wallentowitz1996vaser,chudnovsky2004phonon, li2025dissipation, zhang2018phase, wang2018phonon, lee2023prototype, vashahri2021magnomechanical, Eremeev2020PRA, Eremeev2024AQT, lei2023quantum, khaetskii2013proposal, baur2026qtplarxiv} and realized across a wide range of experimental platforms~\cite{bron1978stimulated, liu2003coupled, vahala2009phonon, mahboob2013phonon, zhang2018phonon, ohtani2019electrically, pettit2019optical, mercade2021floquet, chafatinos2020polariton, zhang2022dissipative, behrle2023phonon, wang2023laser2, xiao2023nonlinear, fu2023photonphonon, kuang2023nonlinear, xiong2023phonon, pan2024closed, knunz2010injection}. Similar to photonic lasers, in phonon lasing the mechanical mode undergoes a threshold-driven transition from incoherent fluctuations to self-sustained coherent macroscopic oscillations. Typically, such macroscopic amplification is achieved through stimulated phonon emission enabled by population inversion in multilevel quantum systems~\cite{bron1978stimulated,vahala2009phonon}, or through dynamical backaction in cavity optomechanical systems~\cite{kippenberg2005analysis,poot2012backaction}. More recently, conventional lasers have been generalized to squeezed lasers, driven-dissipative states that combine the coherence and brightness of conventional lasers with quadrature squeezing~\cite{Navarrete2014,sq_photon_laser2021,Zhao2024PRR,Tian2025,Pontula2025Nanoph}. Above a well-defined lasing threshold, they exhibit self-sustained coherent macroscopic oscillations with reduced fluctuations in one quadrature at the expense of increased fluctuations in its conjugate~\cite{Navarrete2014}. While squeezing and lasing individually underpin numerous advances~\cite{PhysRevX.10.031003,doi:10.1126/science.abb0328,Anderson:17,Mason2019,PhysRevA.53.467,PhysRevA.90.063630,RevModPhys.90.035005,Liang2025NatPhys}, squeezed lasers offer a powerful route to enhancing the sensitivity, spatial and temporal resolution, and range of coherent detection protocols~\cite{Xia2023,Frascella2021,doi:10.1126/science.1160627}. These advances naturally motivate extending the concept of squeezed lasers to the acoustic domain~\cite{lee2026sqarxiv,baur2026qtplarxiv,Behrle2025phlas,Zhang2026,Song2026NatComm}, where self-sustained phonon lasing and quadrature squeezing are unified within a single driven-dissipative state.

Solid-state spin-phonon systems provide a versatile platform for generating and controlling nonclassical mechanical states. In particular, color centers in wide-bandgap materials, such as nitrogen-vacancy (NV) centers in diamond and negatively charged boron vacancies ($V_B^-$) in hexagonal boron nitride (hBN), combine optically addressable electronic spins with strong spin-phonon coupling mediated by strain or magnetic-field gradients~\cite{Bennett2013PRL,Wu2025Nature,Kepesidis2013PRB,PRL_hBN}. In diamond, this interaction enables phonon lasing in the resolved-sideband regime~\cite{Kepesidis2013PRB}. Meanwhile, hBN has emerged as a promising two-dimensional van der Waals platform, where suspended membranes exhibit ultrahigh mechanical quality factors and $V_B^-$ defects possess large strain susceptibility together with room-temperature optically detected magnetic resonance~\cite{Gottscholl2021NatCom,Rizzato2023NatCom,Vaidya2023AdvPhys}. A spin-mechanical scheme based on suspended hBN membranes was recently proposed to generate squeezed mechanical states via ultrastrong far-detuned spin-phonon coupling~\cite{PRL_hBN}. This platform has since been proposed for the generation of nonclassical phonon states and spin-spin entanglement~\cite{Molinares2024EPJQT,Qiao2025PRA}. A further crucial ingredient is a mechanism for engineering the required nonlinear dynamics. To this end, periodic driving has been identified as a powerful route to squeezed lasing in circuit quantum electrodynamics platforms~\cite{Navarrete2014}, while recent optical experiments have realized squeezed lasing using a reservoir-engineered parametric oscillator~\cite{Tian2025}. Yet, most implementations rely on custom-built optical or microwave cavities, posing challenges for on-chip integration. By contrast, solid-state spin-defect systems offer a distinct architectural advantage: Floquet engineering of the spin-phonon interaction provides the effective nonlinear dynamics required for squeezed phonon lasing without optical or microwave cavities.

In this work, we propose a general Floquet-engineered scheme for squeezed phonon lasing that enables a continuous crossover from phase-diffused phonon lasing to phase-locked squeezed phonon lasing. Combining periodic driving~\cite{Bukov2015AdvPhys} with engineered effective spin-mechanical interactions, our scheme generates mechanical gain and quadrature squeezing within a unified framework. As a concrete implementation, we consider color centers embedded in a circular hBN membrane~\cite{PRL_hBN,Tran2016}, where two pairs of periodically driven spins---a primary pair and an ancilla pair---couple to a single mechanical mode and are driven by tailored microwave fields. Under suitable resonance conditions, we show that the primary spins provide controllable mechanical gain, while the ancilla spins engineer Bogoliubov-mode cooling, thereby establishing the steady-state squeezed phonon lasing. Remarkably, the Floquet protocol intrinsically induces phase locking~\cite{PhysRevA.96.013817}, suppressing phase diffusion and thereby stabilizing the steady-state squeezed phonon lasing without external feedback. We thoroughly analyze the steady-state properties of the system, including the lasing threshold, mechanical 
occupation, emission spectrum and second-order correlations. Our work identifies Floquet-engineered spin-phonon systems as a promising route toward nonlinear phononics and the preparation of nonclassical mechanical states in two-dimensional materials~\cite{Gilardoni2025,Fang2024,metrologymaterials2d,Vaidya2023AdvPhys,wang2025quantumsensingdmaterials}.

\section{Floquet Engineering of a Spin-Mechanical System}
\label{sec_floquet}

\subsection{Building-block Hamiltonian}
\label{sec2a}
\begin{figure*}[t]
\centering
\includegraphics[width=1\linewidth]{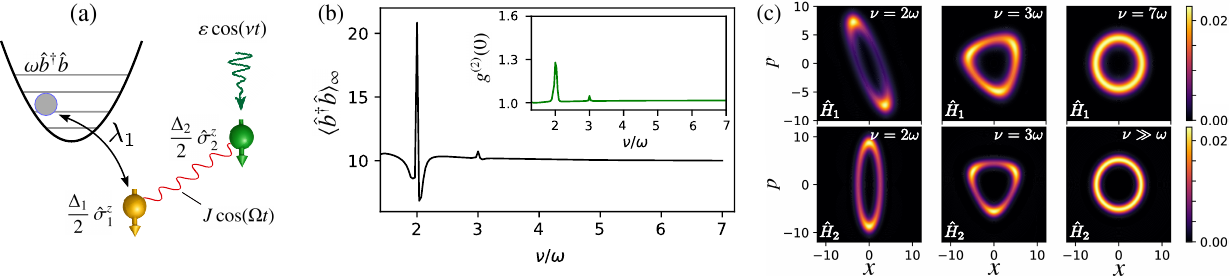}
\caption{
$(a)$ The basic building block of our theoretical proposal for a squeezed phonon laser is a mechanical oscillator (MO) coupled to a spin, which in turn interacts with a second spin via a driven spin-spin coupling; see $\hat{H}_1$ of Eq.~(\ref{eq_hamiltonian}). $\omega$ is the MO frequency whereas $\Delta_1$ and $\Delta_2$ are the energy splitting of the two spins. A local periodic drive is applied to the second spin at frequency $\nu$ and amplitude $\varepsilon$, whereas the spin-spin interaction is modulated at frequency $\Omega$ and amplitude $J$. $(b)$ Steady-state mean phonon number $\langle\hat{b}^{\dagger}\hat{b}\rangle_{\infty}$, by numerically solving Eq. \eqref{ME0} with Hamiltonian $\hat{H}_1$ of Eq.~\eqref{eq_hamiltonian}, as a function of $\nu/\omega$ for the resonance condition $\Omega=\Delta_1+\Delta_2+\omega$. Inset of (b) shows the second-order correlation function $g^{(2)}(0)$ as a function of $\nu/\omega$. $(c)$ Wigner function $W(x,p)$ for selected values of $\nu/\omega$, with $\varepsilon=\nu/2$. 
The top row corresponds to numerical solutions of Eq.~\eqref{ME0} with Hamiltonian $\hat{H}_1$ of Eq.~\eqref{eq_hamiltonian}, while the bottom row shows numerical solutions obtained using the effective Hamiltonian $\hat{H}_2$ of Eq.~\eqref{full_Heff}. The parameters are: $\Delta_1=20\omega$, $\Delta_2=10\omega$, $\lambda_1=0.04\omega$, $J=0.3\omega$, $\Gamma_{1}=\Gamma_{2}=0.03\omega$, $\gamma_m=10^{-3}\omega$, $\bar{n}_{1}^{s}=\bar{n}_{2}^{s}=10^{-5}$, and $\bar{n}_{m}=10^{-3}$.}
\label{fig01}
\end{figure*}

Let us begin by considering a single spin coupled dispersively to a mechanical oscillator (MO)~\cite{treutlein2014hybrid}. A second driven spin is coupled to the first through a time-dependent nearest-neighbor interaction while remaining uncoupled from the MO; see the schematic in Fig.~\ref{fig01}\hyperref[fig01]{(a)}. The total Hamiltonian is ($\hbar = 1$)
\begin{eqnarray}
\label{eq_hamiltonian}
    \hat{H}_{1} 
    &=&\omega \hat{b}^{\dagger}\hat{b}+ \sum_{j=1}^2 \frac{\Delta_j}{2}  \hat{\sigma}_{j}^{z}
    - \lambda_1 \hat{\sigma}_{1}^{z}(\hat{b}^{\dagger}+\hat{b})\nonumber\\
    &+& J \cos(\Omega t)\hat{\sigma}_{1}^{x}\hat{\sigma}_{2}^{x}
    + \varepsilon \cos(\nu t) \hat{\sigma}_{2}^{z},
\end{eqnarray}
where $\hat{\sigma}_j^{\alpha}$, with $\alpha = x,y,z$, denote the Pauli operators acting on the spin at site $j$ along the $\alpha$ direction and $\hat{b}$ ($\hat{b}^\dagger$) denote the annihilation (creation) operators of the MO. The parameter $\Delta_j$ specifies the energy splitting of the spin system, while $\omega$ is the frequency of the MO. The spin-mechanical coupling strength is denoted by $\lambda_1$ and can be engineered using different experimental platforms~\cite{treutlein2014hybrid}. The parameter $J$ denotes the amplitude of the external driving field, while $\Omega$ is its driving frequency. A local drive with amplitude $\varepsilon$ and frequency $\nu$ is applied to the second spin, enabling Floquet engineering \cite{pena2025steering}.

\subsection{Effective Floquet Hamiltonian for two spins}
\label{sec2b}
To gain insight into the mechanism underlying the mechanical amplification, we derive an effective Hamiltonian from Eq.~\eqref{eq_hamiltonian} by means of a Floquet decomposition in terms of the harmonics of the driving frequency $\nu$. To this end, we perform an appropriate transformation to a rotating frame, apply the rotating-wave approximation under the corresponding resonance conditions, scale the Hamiltonian in Eq.~\eqref{eq_hamiltonian} by $\omega$, and assume the weak-coupling regime between the spin and the MO. This procedure leads to an effective Hamiltonian that explicitly depends on the local driving frequency $\nu$ (see details in Appendix \ref{sect_S1}), given by:
\begin{eqnarray}\label{full_Heff}
\hat{H}_2\left(\frac{\nu}{\omega}\right)
&=&J_{\text{eff}}\hat{\sigma}_1^{+} \hat{\sigma}_2^{+}+i\tilde{g}_{\text{eff}}\Bigg(\left[\mathcal{J}_{\beta'}(\chi){-}\mathcal{J}_{\alpha'}(\chi)\right]{b}^{\dagger}\nonumber\\
    &+&\left[\mathcal{J}_{\delta'}(\chi){-}\mathcal{J}_{\alpha'}(\chi)\right]\hat{b}\Bigg)\hat{\sigma}_1^{+} \hat{\sigma}_2^{+} + \mathcal{O}\left(\epsilon^2\right){+}\text{H.c.}
\end{eqnarray}
In the above, we have defined the parameters
\begin{eqnarray}
\chi &=& \frac{2\varepsilon}{\nu},\\
\epsilon &=& \frac{\lambda_1}{\omega},\\
\tilde{g}_{\mathrm{eff}} &=& \frac{J\lambda_1}{\omega},\\
J_{\mathrm{eff}} &=& \frac{J}{2}\mathcal{J}_{\alpha'}(\chi),
\label{eq_bessel}
\end{eqnarray}
The effective Hamiltonian of Eq.~\eqref{full_Heff} is derived under the resonance condition
\begin{equation} \label{res_ampl}
\Omega=\Delta_1+\Delta_2 + \omega,
\end{equation}
for which the Bessel function $\mathcal{J}_{m}(z)$ of the first kind of order $m=\alpha',\beta',\delta'$, gets the following indices $\alpha'=\omega/\nu$, $\beta'=0$, and $\delta'=2\omega/\nu$, with $\alpha',\delta'\in\mathbb{Z}$; see Appendix~\ref{sect_S1} (Eqs. \ref{coef_u} -- \ref{coef_v}) for details. Under these conditions, the effective Hamiltonian describes the coherent coupling between a pair of spin-flip operators and the bosonic mode. The interaction is selectively activated by the resonance conditions and has same structure as that of the phonon-laser model proposed in~\cite{phonon_lasers2026}.

\subsection{Squeezed phonon amplification with two spins}
\label{subsect:Squeezing-amplification}

Lasing is an inherently nonequilibrium phenomenon that emerges in driven-dissipative quantum systems. To describe this dynamics, we employ a Markovian quantum master equation (ME) in Lindblad form~\cite{petruccione2002theory}. Accordingly, the density operator $\hat{\rho}$ of the coupled spin-mechanical system evolves as
\begin{eqnarray}\label{ME0}
     \frac{d\hat{\rho}}{dt}&=&-i[ \hat{H}, \hat{\rho}]+\gamma_m\left(1+\bar{n}_{m}\right)\mathcal{L}_{\hat{b}}[\hat{\rho}]+\gamma_m\bar{n}_{m}\mathcal{L}_{\hat{b}^\dagger}[\hat{\rho}]\nonumber\\
     &+&\sum_{j=1}^2 \left\{\Gamma_{j}\left(1+\bar{n}_{j}^{s}\right)\mathcal{L}_{\hat{\sigma}_{j}^{-}}[\hat{\rho}]+\Gamma_{j}\bar{n}_{j}^{s}\mathcal{L}_{\hat{\sigma}_{j}^{+}}[\hat{\rho}]
     \right\},
\end{eqnarray}
where $\hat{H}$ is a general Hamiltonian. Here the notation $\mathcal{L}_{\hat{\mathcal{O}}}[\hat{\rho}]$ stands for the dissipation superoperator of a decay channel with the associated jump operator $\hat{O}$; specifically $\mathcal{L}_{\hat{\mathcal{O}}}[\hat{\rho}] = \hat{\mathcal{O}}\hat{\rho} \hat{\mathcal{O}}^{\dagger}-\big (\hat{\mathcal{O}}^{\dagger}\hat{\mathcal{O}}\hat{\rho}+\hat{\rho}\hat{\mathcal{O}}^{\dagger}\hat{\mathcal{O}}\big)/2$, $\Gamma_{j}$ is the decay rate of each spin $j$, and $\gamma_{m}$ is the damping of the mechanical mode, with $\bar{n}_{j}^{s}$ ($\bar{n}_{m}$) denoting the mean occupation number of the thermal reservoir coupled to the $j$th spin (mechanical mode).

In the following, we numerically solve the ME~\eqref{ME0} using the Hamiltonians of Eq.~\eqref{eq_hamiltonian} and Eq.~\eqref{full_Heff}. Unless otherwise stated, the spin system is initialized in the product state $\lvert \psi_0 \rangle = |{\downarrow}\rangle |{\downarrow}\rangle$, while the MO is prepared in a thermal state with mean occupation number $\bar{n}_{m}=10^{-3}$.

In Fig.~\ref{fig01}\hyperref[fig01]{(b)}, we show the steady-state phonon occupation $\langle \hat{b}^\dagger \hat{b}\rangle_{\infty}$ as a function of the local driving frequency $\nu/\omega$, with the spin-spin driving frequency fixed at the resonance condition ~\eqref{res_ampl}, i.e. $\Omega=\sum_{j=1}^{2}\Delta_j+\omega$. As the figure shows, a pronounced twofold mechanical amplification occurs at $\nu=2\omega$. The inset of Fig.~\ref{fig01}\hyperref[fig01]{(b)} shows the steady-state zero-delay second-order correlation function $g^{(2)}(0)$ as a function of $\nu/\omega$. As the figure shows, $g^{(2)}(0)$ gradually approaches the coherent limit $g^{(2)}(0)\approx 1$ as $\nu/\omega$ increases. In contrast, a pronounced peak, reaching $g^{(2)}(0)\approx 1.3$, emerges at $\nu=2\omega$, revealing enhanced phonon-number fluctuations at the mechanical amplification point. Note that such an enhancement of $g^{(2)}(0)$ is characteristic of squeezed lasing, for which the steady-state value converges to $g^{(2)}(0)=3/2$~\cite{sq_photon_laser2021}.

In Fig.~\ref{fig01}\hyperref[fig01]{(c)}, we present the steady-state Wigner quasiprobability distribution,
\begin{equation}
W(x,p)=\frac{1}{\pi\hbar}\int_{-\infty}^{\infty}\langle x+y|\hat{\rho}_{\mathrm{MO}}|x-y\rangle e^{-2ipy/\hbar}dy,
\end{equation}
of the reduced MO state for different values of $\nu/\omega$. The Wigner function is represented in the quadrature phase space defined by the dimensionless operators $\hat{x}=(\hat{b}+\hat{b}^{\dagger})/2$ and $\hat{p}=(\hat{b}-\hat{b}^{\dagger})/(2i)$. The bottom row of Fig.~\ref{fig01}\hyperref[fig01]{(c)} shows the steady-state Wigner functions obtained from the effective Hamiltonian $\hat{H}_2(\nu/\omega)$ in Eq.~\eqref{full_Heff}, expressed in the rotating frame. By contrast, the top row displays the corresponding Wigner functions obtained by numerically solving the ME~\eqref{ME0} in the laboratory frame. In the latter case, the Wigner function rotates in phase space at the mechanical frequency $\omega$, and we therefore present a representative snapshot. Overall, the effective description is in excellent agreement with the full numerical solution.

Interestingly, the Wigner distributions in the leftmost column of Fig.~\ref{fig01}\hyperref[fig01]{(c)} reveal that, at the resonance $\nu=2\omega$, the MO undergoes amplification of a squeezed mode~\cite{sq_photon_laser2021}. To uncover the origin of this behavior, we evaluate the effective Hamiltonian in Eq.~\eqref{full_Heff} at $\nu=2\omega$. We introduce the coefficients
\begin{equation} \label{def_uv}
\mathrm{u}=\frac{\mathcal{J}_{0}(\chi)}{N},
\qquad
\mathrm{v}=\frac{\mathcal{J}_{1}(\chi)}{N},
\end{equation}
where
\begin{equation}
N=\sqrt{\left|\mathcal{J}_{0}^{2}(\chi)-\mathcal{J}_{1}^{2}(\chi)\right|},
\end{equation}
such that the Bogoliubov coefficients satisfy the canonical relation
\begin{equation}
|\mathrm{u}|^{2}-|\mathrm{v}|^{2}=1.
\end{equation}
Under the condition $|\mathrm{u}|>|\mathrm{v}|$, the effective Hamiltonian reads (see Fig. \ref{figS3}\hyperref[figS3]{(a)} in Appendix ~\ref{sm4})
\begin{eqnarray}
\hat{H}_2\left(\frac{\nu}{\omega}=2\right)
&=&
ig_{\mathrm{eff}}
\left(
\mathrm{u}\hat{b}^{\dagger}
+\mathrm{v}\hat{b}
\right)
\hat{\sigma}_{1}^{+}\hat{\sigma}_{2}^{+}
+\text{H.c.}\nonumber\\
&\equiv&
ig_{\mathrm{eff}}
\hat{B}^{\dagger}
\hat{\sigma}_{1}^{+}\hat{\sigma}_{2}^{+}
+\text{H.c.},
\label{eq:heff1_l}
\end{eqnarray}
where $g_{\mathrm{eff}}=\tilde{g}_{\mathrm{eff}}N$ is the renormalized coupling strength and
\begin{equation}\label{B_op}
\hat{B}=\mathrm{u}\hat{b}+\mathrm{v}\hat{b}^{\dagger}
\end{equation}
is a Bogoliubov (squeezed) bosonic mode obeying the canonical commutation relation $[\hat{B},\hat{B}^{\dagger}]=1$.

Eq.~\eqref{eq:heff1_l} shows that the effective interaction coherently couples the simultaneous excitation of the two spins to the creation of a squeezed mode excitation $\hat{B}$. Consequently, the driven-dissipative dynamics selectively amplifies the Bogoliubov mode $\hat{B}$ rather than the bare mechanical mode $\hat{b}$. In the resonant regime, all nonresonant Floquet components are eliminated within the rotating-wave approximation, leaving a phase-sensitive interaction that mediates two-spin excitation processes through the squeezed mode. Throughout the numerical simulations shown in Fig.~\ref{fig01}\hyperref[fig01]{(b)-(c)}, we set $\chi=1$ (i.e., $\nu = 2\varepsilon$), for which $|\mathrm{u}|>|\mathrm{v}|$ (see details in Appendix \ref{sm4} together with a summary of feasible experimental parameters used in our work).

To understand the behavior at the resonance $\nu=3\omega$, shown in the middle column of Fig.~\ref{fig01}\hyperref[fig01]{(c)}, we evaluate the effective Hamiltonian in Eq.~\eqref{full_Heff} at $\nu = 3\omega$. In this regime, the effective interaction contains both single-phonon and two-phonon processes. The corresponding effective Hamiltonian reads
\begin{eqnarray}
\hat{H}_2\left(\frac{\nu}{\omega}=3\right)
&=&
i\tilde{g}_{\mathrm{eff}}\mathcal{J}_{0}(\chi)
\hat{b}^{\dagger}
\hat{\sigma}_{1}^{+}\hat{\sigma}_{2}^{+}
\nonumber\\
&-&
J\left(\frac{\lambda_{1}}{\omega}\right)^{2}
\mathcal{J}_{1}(\chi)
\hat{b}^{2}
\hat{\sigma}_{1}^{+}\hat{\sigma}_{2}^{+}
+\text{H.c.}
\end{eqnarray}
The coexistence of these competing excitation channels is responsible for the characteristic phase-space structure observed in the Wigner function~\cite{three_mode_photon}. By contrast, for sufficiently large driving frequencies, the system crosses over to the conventional phonon-lasing regime, as illustrated by the rightmost column of Fig.~\ref{fig01}\hyperref[fig01]{(c)} for $\nu = 7\omega$. This behavior is expected because the Floquet-induced multiphonon processes become progressively weaker as $\nu/\omega$ increases. Indeed, in the limit $\nu\gg\omega$, the effective Hamiltonian reduces to
\begin{equation}
\hat{H}_2(\nu\gg\omega)
\simeq
i\tilde{g}_{\mathrm{eff}}
\mathcal{J}_{0}(\chi)
\hat{b}^{\dagger}
\hat{\sigma}_{1}^{+}\hat{\sigma}_{2}^{+}
+\text{H.c.},
\end{equation}
which coincides with the effective Hamiltonian obtained in the model proposed in ~\cite{phonon_lasers2026}, where it gives rise to conventional phonon lasing.

Before concluding this section, it is worth emphasizing that the results presented in Fig.~\ref{fig01} demonstrate that the system exhibits both mechanical amplification $\langle \hat{b}^\dagger \hat{b} \rangle_\infty \sim 20$ and quadrature squeezing for $\nu = 2\omega$. Nevertheless, the generated state is a statistical mixture of squeezed coherent states~\cite{Navarrete2014}. This is reflected in the Wigner function, whose height is distributed unevenly around the squeezed ring in phase space, indicating partial phase localization rather than the uniform phase distribution characteristic of a fully phase-diffused lasing state. The reason is that the mechanical dissipation acts on the bare phonon mode $\hat{b}$ rather than on the Bogoliubov mode $\hat{B}$. Consequently, coherent amplification of the squeezed mode competes with losses in the original mechanical basis, preventing the emergence of a steady-state squeezed phonon laser. In the next section, we show that this limitation can be overcome by extending the system with an additional pair of spins and operating under a different resonance condition, thereby enabling a fully developed steady-state squeezed phonon laser. 

\section{Squeezed Phonon Laser in a Hexagonal Boron Nitride ({\protect\MakeLowercase{h}}BN) Membrane with Color Centers}
\label{sec_hBN}

\subsection{Total spin-mechanical Hamiltonian in hBN}
\label{sec_hBNa}

To establish a steady-state squeezed phonon laser, coherent amplification of the Bogoliubov mode must be accompanied by engineered dissipation in the same squeezed basis, i.e., squeezed cooling~\cite{Navarrete2014}. We achieve this by introducing an ancillary pair of spins that mirrors the primary spin pair and is governed by the same spin-mechanical interaction, while operating under a different resonance condition. We also discuss a feasible implementation of the proposed scheme using color-center spins embedded in a suspended hBN mechanical membrane.

\begin{figure}[t]
\centering
\includegraphics[width=0.92\linewidth]{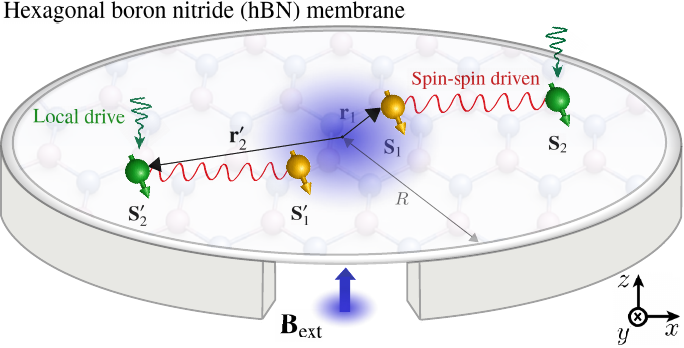}
\caption{Sketch of the proposed squeezed phonon laser based on a suspended hexagonal boron nitride (hBN) membrane. A suspended circular hBN membrane of radius $R$ acts as the mechanical resonator. A magnetic field gradient enables spin-motion coupling with spins placed at $\mathbf{r}_{1}(\mathbf{r}'_{1})$ (yellow spheres), each interacting with a second spin placed at $\mathbf{r}_{2}(\mathbf{r}'_{2})$ (green spheres) via a driven spin-spin coupling. A local periodic drive is applied to the second spins.}
\label{fighBN}
\end{figure}

Let us begin by considering four spin qubits embedded in a suspended hBN mechanical membrane (see Fig.~\ref{fighBN}): two primary spin qubits $\mathbf{S}_1$ and $\mathbf{S}_2$ located at positions $\mathbf{r}_{1}$ and $\mathbf{r}_{2}$, respectively; and two ancillary spin qubits $\mathbf{S}'_1$ and $\mathbf{S}'_2$ located at positions $\mathbf{r}'_{1}$ and $\mathbf{r}'_{2}$, all subject to an external magnetic field $\mathbf{B}_{\mathrm{ext}}$. The coherent spin dynamics is described by the Zeeman Hamiltonian $\hat{H}_{SB}= \mu_B g_e\sum_{j=1}^{2}\left(\mathbf{S}_j\cdot\mathbf{B}_{\mathrm{ext}}+\mathbf{S}'_j\cdot\mathbf{B}_{\mathrm{ext}}\right).$ Without loss of generality, we assume a uniform magnetic field oriented along the $z$ direction, $\mathbf{B}_{\mathrm{ext}}=B_{z}\hat{z}\equiv B_{\perp}$, where $\mu_B$ and $g_e$ denote the Bohr magneton and the electron gyromagnetic ratio, respectively. Expanding the Zeeman Hamiltonian about the mechanical equilibrium position and promoting the mechanical displacement to a quantum operator yields the following Hamiltonian governing the coherent dynamics of the four-spin system~\cite{PRL_hBN}\footnote{We note that the Hamiltonian derived in Ref.~\cite{PRL_hBN} contains an additional term associated with a magnetic field parallel to the membrane. We omit this term here because our analysis shows that, under experimentally relevant conditions, its contribution to the dynamics is negligible. A detailed justification is provided in Section~\ref{sec_hBN_apx}}:
\begin{eqnarray}\label{eq_membrane}
    \hat{H}_{S}&\approx& \sum_{j=1}^{2}\left (\frac{\Delta_j}{2} \hat{\sigma}_{j}^{z}
    +\frac{\Delta'_j}{2} \hat{\sigma}_{j}'^{z}\right)
    +\omega\hat{b}^{\dagger}\hat{b}\nonumber\\
    &-&\lambda_1\hat{\sigma}^{z}_{1}(\hat{b}+\hat{b}^{\dagger})
    -\lambda'_1\hat{\sigma}_{1}'^{z}(\hat{b}+\hat{b}^{\dagger}).
\end{eqnarray}
The parameter $\Delta_j\equiv\Delta(r_j,\theta_j)=\mu_B g_e B_{\perp}(r_j,\theta_j,0)$ denotes the spin-qubit energy splitting, while $\lambda_j \equiv \lambda(r_j,\theta_j) = \mu_B g_e 
\partial_z B_{\perp}(r_j,\theta_j,0) \Psi(r_j,\theta_j) \sqrt{\hbar/(2m\omega)}$
is the spin-mechanical coupling strength. Here, $\Psi(r_j,\theta_j)$ is the mechanical mode profile evaluated at the defect position, and $\sqrt{\hbar/(2m\omega)}$ is the amplitude of the zero-point motion of the mechanical mode, with $m$ and $\omega$ denoting its effective mass and resonance frequency, respectively~\cite{PRL_hBN}. Analogous definitions apply to the ancillary spin qubits.

We assume that the spin-mechanical couplings associated with the spin operators $\hat{\sigma}^{z}_{2}$ and $\hat{\sigma}'^{z}_{2}$ are negligible, i.e., $(\lambda_{2},\lambda'_{2})\ll1$. This condition is fulfilled when the corresponding spin qubits are positioned close to a node of the fundamental mechanical mode, where the mode profile satisfies $\Psi=0$. For a circular membrane with the boundary conditions considered in Ref.~\cite{PRL_hBN}, this corresponds approximately to $\mathbf{r}_{2}/R\sim1$ and $\mathbf{r}'_{2}/R\sim1$, where $R$ is the membrane radius. Typical suspended hBN membranes have radii of a few $\mu\mathrm{m}$ and mechanical resonance frequencies on the order of hundreds of MHz (see details in Appendix \ref{sm4}).

To realize a phonon laser mediated by coherent spin-spin interactions, a robust and controllable exchange interaction between the spin qubits is essential~\cite{phonon_lasers2026}. To this end, we consider a system in which the exchange interaction within each spin pair is periodically modulated by an external field~\cite{PRB_Mod,PRB_Mod2} with period $T=2\pi/\Omega$. The corresponding time-dependent Hamiltonian is
\begin{equation}
\hat{H}_{J}
=
J\cos(\Omega t)\mathbf{S}_1\!\cdot\!\mathbf{S}_2
+
J'\cos(\Omega' t)\mathbf{S}'_1\!\cdot\!\mathbf{S}'_2,
\end{equation}
where $\Omega$ ($\Omega'$) and $J$ ($J'$) denote the modulation frequency and amplitude of the exchange interaction for the primary (ancillary) spin pair, respectively. In the absence of an external electric field, exchange couplings on the order of MHz can be achieved for spin-pair separations of a few $\mu\mathrm{m}$ (see details in Appendix \ref{sm4}).

Without loss of generality, we restrict the exchange interaction to its $xx$ component, in which case
\begin{equation}
\label{eq_spin-spin_hamiltonian}
\hat{H}_{J}
=
J\cos(\Omega t)\hat{\sigma}_{1}^{x}\hat{\sigma}_{2}^{x}
+
J'\cos(\Omega' t)\hat{\sigma}_{1}'^{x}\hat{\sigma}_{2}'^{x}.
\end{equation}
This interaction constitutes the second key ingredient for realizing a steady-state squeezed phonon laser.


The realization of a squeezed phonon laser also requires local spin driving. In principle, independent time-periodic electric fields can be applied to individual spin qubits. In practice, however, the close spatial proximity of neighboring qubits may lead to cross-talk arising from the finite spatial extent of the driving fields~\cite{cross-talk}. To mitigate this effect, we apply local driving only to the spin qubits located at the extremities of the membrane (indicated by the green spheres in Fig.~\ref{fighBN}), where their larger spatial separation naturally suppresses cross-talk. The corresponding Hamiltonian is
\begin{equation}
\label{eq_spin-local_hamiltonian}
\hat{H}_{LD}
=
\varepsilon \cos(\nu t)\hat{\sigma}_{2}^{z}
+
\varepsilon' \cos(\nu' t)\hat{\sigma}_{2}'^{z},
\end{equation}
where $\varepsilon$ ($\varepsilon'$) and $\nu$ ($\nu'$) denote the amplitude and frequency of the local driving field acting on the primary (ancillary) spin qubit, respectively. Eq.~\eqref{eq_spin-local_hamiltonian} provides the time-periodic driving required to engineer the effective Floquet interactions discussed in Sec.\ref{sec2b}.


We now have all the ingredients required to construct the full Hamiltonian leading to squeezed phonon lasing: the coherent four-spin Hamiltonian $\hat{H}_{S}$ in Eq.~\eqref{eq_membrane}, the spin-spin exchange interaction $\hat{H}_{J}$ in Eq.~\eqref{eq_spin-spin_hamiltonian}, and the local spin driving $\hat{H}_{LD}$ in Eq.~\eqref{eq_spin-local_hamiltonian}. The total hBN Hamiltonian is therefore $\hat{H}_\mathrm{hBN}=\hat{H}_{S}+\hat{H}_{J}+\hat{H}_{LD}$. To make the respective roles of amplification and cooling explicit, we rewrite the total Hamiltonian as
\begin{eqnarray}
\label{eq_Htot}
\nonumber\hat{H}_\mathrm{hBN}&=&\hat{H}_{S}+\hat{H}_{J}+\hat{H}_{LD}\\
&\equiv&\omega\hat{b}^{\dagger}\hat{b}+\hat{H}^{A}+\hat{H}^{C},
\end{eqnarray}
where
\begin{eqnarray}
    \nonumber\hat{H}^{A}&=&\sum_{j=1}^{2}\frac{\Delta_j}{2} \hat{\sigma}_{j}^{z}-\lambda_1\hat{\sigma}_{1}^{z}(\hat{b}^{\dagger}+\hat{b})\\
    &+&J\cos{(\Omega t)}\hat{\sigma}_{1}^{x}\hat{\sigma}_{2}^{x}+\varepsilon\cos{(\nu t)} \hat{\sigma}_{2}^{z},
\end{eqnarray}
denotes the amplification Hamiltonian and 
\begin{eqnarray}
    \nonumber\hat{H}^{C}&=&\sum_{j=1}^{2}\frac{\Delta_j'}{2} \hat{\sigma}_{j}'^{z}-\lambda_1'\hat{\sigma}_{1}'^{z}(\hat{b}^{\dagger}+\hat{b})\\
    &+&J'\cos{(\Omega' t)}\hat{\sigma}_{1}'^{x}\hat{\sigma}_{2}'^{x}+\varepsilon'\cos{(\nu' t)} \hat{\sigma}_{2}'^{z}.\label{eq_Hcooling}
\end{eqnarray}
denotes the cooling Hamiltonian, respectively. The primary spin qubits (described by the operators $\hat{\sigma}_{j}$) mediate the squeezed amplification process under the targeted frequency resonance conditions discussed in the previous section, whereas the ancillary spin qubits (described by $\hat{\sigma}'_{j}$) implement the squeezed cooling process under different resonance conditions, as discussed below. Note that we also consider the effect of adding a static driving term $\frac{\Lambda_j}{2}\hat{\sigma}_j^x$ to the Hamiltonian in Eq.~\eqref{eq_Htot}, where $\Lambda_j$ denotes the Rabi frequency at site $j$ (see Appendix~\ref{sec_hBN_apx}). We emphasize that, within the parameter regime considered throughout this work, this additional term does not contribute to the effective Hamiltonian $\hat{H}_{\mathrm{hBN}}$. It can therefore be safely neglected in the main analysis.

\subsection{Squeezed phonon lasing stabilized by ancilla spins}\label{sec4}

To demonstrate the emergence of stable squeezed phonon lasing, we use the ancillary spin qubits to engineer phonon dissipation through the squeezed-cooling Hamiltonian $\hat{H}^{C}$ in Eq.~\eqref{eq_Hcooling}~\cite{Porras2012,Navarrete2014}. For this analysis, the auxiliary qubits are controlled by the same parameters as described in the previous section, with the critical exception of setting the resonance condition
\begin{equation}
\Omega'=\Delta'_1 + \Delta'_2-\omega.
\end{equation}
This condition simply interchanges the values of $\mathrm{u}$ and $\mathrm{v}$, as defined in Eq.~\eqref{def_uv}, thereby leading directly to the following effective Hamiltonian:
\begin{eqnarray}
\hat{H}_\text{eff}^{C}&=&ig'_{\mathrm{eff}}
\left(\mathrm{u}\hat{b}+\mathrm{v}\hat{b}^{\dagger}\right) \hat{\sigma}_1'^{+}\hat{\sigma}_2'^{+} +\text{H.c.}\nonumber\\
&\equiv& ig'_{\text{eff}}\hat{B} \hat{\sigma}_1'^{+} \hat{\sigma}_2'^{+}+\text{H.c.},
\end{eqnarray}
where the $\hat{B}$ is defined in Eq. \eqref{B_op}, $\hat{\sigma}_1'^{+}$, $\hat{\sigma}_2'^{+}$ correspond to operators of the auxiliary qubits, and $g'_{\text{eff}} $ is the effective spin-mechanical coupling strength.

To achieve squeezed phonon losses in the ME, we perform an adiabatic elimination in Eq. \eqref{eq_Htot}, assuming a fast decay of the ancilla qubits \cite{petruccione2002theory, Porras2012}. For the adiabatic approximation to hold, the condition $g'_{\mathrm{eff}} \ll \Gamma'_j$ must be satisfied, where $g'_{\mathrm{eff}} = NJ'\lambda'_1/\omega$. Since the spin-mechanical coupling $\lambda'_1$ is fixed by the membrane geometry and the magnetic field configuration, it cannot be tuned independently. Therefore, the adiabatic approximation can be enforced by controlling the amplitude of the external driving field, such that $J'\ll\Gamma'_j$. As a result, one obtains an effective ME describing the dynamics of the full system defined in Sec.~\ref{sec_hBNa}:
\begin{eqnarray}\label{ME2}
     \frac{d\hat{\rho}}{dt}&=&-i[\hat{H}_{\text{eff}}^{A},\hat{\rho}] +\gamma'_m\mathcal{L}_{\hat{B}}[\hat{\rho}] \nonumber\\
     &+&\gamma_m\left(1+\bar{n}_{m}\right)\mathcal{L}_{\hat{b}}[\hat{\rho}]+\gamma_m\bar{n}_{m}\mathcal{L}_{\hat{b}^\dagger}[\hat{\rho}]\nonumber\\
     &+&\sum_{j=1}^2\left\{ \Gamma_{j} \left(1+\bar{n}_{j}^{s}\right)\mathcal{L}_{\hat{\sigma}_{j}^{-}}[\hat{\rho}]+\Gamma_{j}\bar{n}_{j}^{s}\mathcal{L}_{\hat{\sigma}_{j}^{+}}[\hat{\rho}]\right\},
\end{eqnarray}
where $\hat{H}_{\mathrm{eff}}^{A}\equiv\hat{H}_2(\nu/\omega=2)$ is defined in Eq. \eqref{eq:heff1_l}. Here, the damping of the squeezed mode $\hat{B}$ is characterized by the decay rate $\gamma'_m \sim g_{\mathrm{eff}}'^2/\Gamma'_{\mathrm{eff}}$, where $\Gamma'_{\mathrm{eff}}$ denotes the effective decay rate of the ancilla spins. To realize the squeezed phonon laser, we consider the regime $\gamma'_m \gg \gamma_m$, where the effective dissipator $\mathcal{L}_{\hat{B}}[\hat{\rho}]$ dominates over the intrinsic phonon dissipation $\mathcal{L}_{\hat{b}}[\hat{\rho}]$.

\begin{figure}[t]
\centering
\includegraphics[width=1\linewidth]{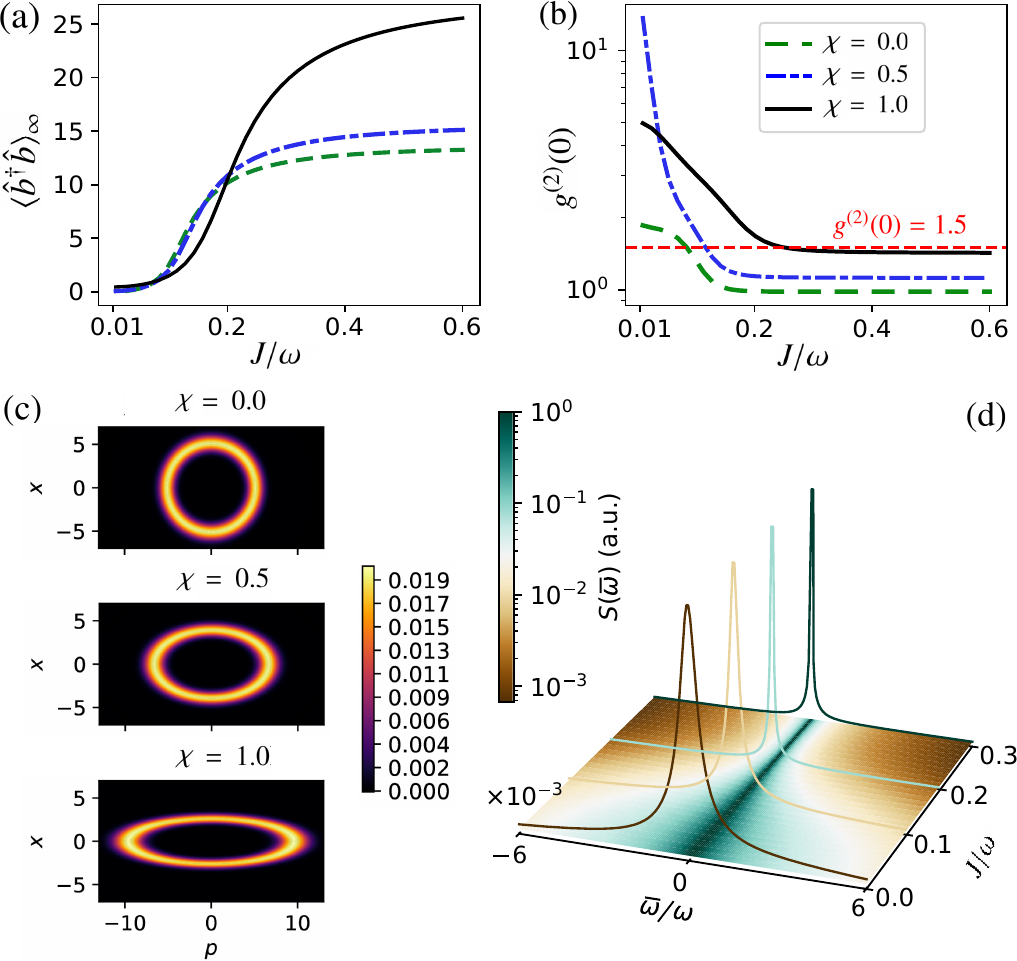}
\caption{All panels show the steady-state solution of the ME in Eq.~\eqref{ME2}. $(a)$ Mean phonon number $\langle\hat{b}^{\dagger}\hat{b}\rangle_{\infty}$ and $(b)$ second-order correlation function $g^{(2)}(0)$ as functions of the driving amplitude $J/\omega$ for different values of $\chi$. $(c)$ Wigner representation $W(x,p)$ at $J=0.3\omega$ for three different values of $\chi$. $(d)$ Power spectrum $S(\overline{\omega})$, in arbitrary units (a.u.), as a function of $J/\omega$ for $\chi=1$. Other parameters are $\lambda_1=0.04\omega$, $\nu=\nu'=2\omega$, $\Gamma_{j}=0.03\omega$, $\gamma_m=10^{-4}\omega$, $\gamma'_m=10\gamma_m$, $\bar{n}_{j}^{s}=10^{-5}$, and $\bar{n}_{m}=10^{-3}$.}
\label{fig3}
\end{figure}

In Fig.~\ref{fig3}\hyperref[fig3]{(a)}, we show the steady-state mean phonon number $\langle \hat{b}^{\dagger}\hat{b}\rangle$, obtained by numerically solving the ME~\eqref{ME2}, as a function of $J/\omega$ for several values of $\chi$. A pronounced buildup of the steady-state phonon population is observed for $J \gtrsim 0.2\omega$ at $\chi=1$. This threshold behavior marks the mechanical analogue of a lasing transition, where Floquet-assisted inter-spin interactions provide the effective pumping that drives phonon amplification beyond a critical drive strength. 

The nature of the lasing state is further characterized by the zero-delay second-order correlation function $g^{(2)}(0)$ shown in Fig.~\ref{fig3}\hyperref[fig3]{(b)}. Without local spin driving $\chi=0$, $g^{(2)}(0)\rightarrow 1$ deep in the lasing regime, indicating coherent, Poissonian phonon statistics characteristic of a conventional laser. A qualitatively different regime emerges when local spin driving is activated and $\chi=1$ is chosen. For this case, $g^{(2)}(0) \approx 1.5$ (red dashed line in Fig.~\ref{fig3}\hyperref[fig3]{(b)}), significantly exceeding the Poissonian value of unity and saturating near the analytical prediction of $3/2$ for squeezed lasing modes~\cite{sq_photon_laser2021}. Similar super-Poissonian statistics have also been reported for squeezed states in other physical platforms~\cite{kam26superpoissonsqueeze,tasci2025superpoissoniansqueezed}.

In Fig.~\ref{fig3}\hyperref[fig3]{(c)}, we show the corresponding steady-state Wigner functions for several values of $\chi$. The first row of Fig.~\ref{fig3}\hyperref[fig3]{(c)} exhibits the characteristic doughnut-shaped annular distribution, a direct signature of a phase-diffused coherent lasing state, consistent with the resonance condition $\Omega=\sum_{j=1}^{2}\Delta_j+\omega$~\cite{phonon_lasers2026}. By contrast, the bottom row demonstrates the emergence of a squeezed phonon laser, as the Wigner function is clearly compressed along the $x$ quadrature and broadened along the conjugate $p$ quadrature. These results show that the degree of squeezing can be continuously controlled through the local driving parameter $\chi$.

Finally, a genuine lasing regime must exhibit linewidth narrowing. To investigate this, we analyze the spectral properties of the emitted phonon field by studying its steady-state power spectrum $S(\overline{\omega})\equiv\int_{-\infty}^{+\infty}d\tau e^{-i\overline{\omega}\tau}
\langle\hat{b}^{\dagger}(\tau)\hat{b}(0)\rangle_{\infty}$, where the expectation value is evaluated with respect to the steady state of the effective Hamiltonian $\hat{H}_2$ in Eq.~\eqref{full_Heff}. In Fig.~\ref{fig3}\hyperref[fig3]{(d)}, we plot $S(\overline{\omega})$ as functions of $\overline{\omega}$ and $J$ for the amplification resonance $\nu=2\omega$. A progressive and pronounced narrowing of the emission linewidth is observed as $J$ increases beyond the lasing threshold, reflecting the buildup of long-lived temporal coherence in the phonon field. This linewidth narrowing is a universal spectral signature of lasing---the mechanical analogue of the Schawlow--Townes linewidth narrowing in optical lasers \cite{Schawlow1958}. Its persistence in the squeezed lasing regime demonstrates that the squeezed phonon laser retains the spectral coherence characteristic of conventional lasing while simultaneously exhibiting the nonclassical correlations associated with a squeezed state, see some discussion in ~\cite{sq_photon_laser2021}.

Note that our proposed squeezed phonon laser can be continuously tuned \emph{in situ} through the spin-driving parameters, namely the modulation amplitudes $\varepsilon$, inter-spin couplings $J$, and driving frequencies $\nu$ and $\Omega$. This high degree of tunability enables dynamic control of the lasing operation, including switching between conventional and squeezed phonon lasing within the same device.

\section{Spin-Controlled Phase-Locked Squeezed Phonon Lasing}
\label{sec_phase_control}

Conventional lasers exhibit phase diffusion. However, the introduction of a weak coherent driving field explicitly breaks the phase symmetry, thereby selecting a preferred phase and leading to phase locking~\cite{PhysRevA.96.013817}, as discussed in Appendix \ref{sm3}. Phase-locked lasing has been shown to play a key role in the self-organized synchronization of phonon-laser arrays~\cite{phonon_lasers2026}. In this section, we propose a scheme in which phase-locked squeezed phonon lasing is achieved without introducing an additional coherent drive to the original system. Instead, we exploit a controlled time-dependent spin-spin interaction that induces correlated spin dynamics, thereby stabilizing the phase of the emitted phonon field and enabling phase-locked squeezed phonon lasing.

\subsection{Primary spin pair at resonance $\nu=\omega$}

To achieve phase-locked squeezed amplification, we now impose the resonance condition as in Eq. \eqref{res_ampl} on primary spins by setting $\nu=\omega$ in Eq. \eqref{full_Heff}. Therefore, we obtain the following orders for the Bessel function: $\alpha'=\omega/\nu=1$, $\beta'=0$, and $\delta'=2\omega/\nu=2$, see the definitions in Appendix~\ref{sect_S1} (Eqs. \ref{coef_u} -- \ref{coef_v}). Under these conditions, one obtains the effective Hamiltonian
\begin{equation} 
\label{eq:heff2_ll}
\hat{H}_\text{locked}^{A}=i g_{1,\text{eff}}
\hat{B}_\text{locked}^{\dagger}\hat{\sigma}_1^{+} \hat{\sigma}_2^{+}
+
\frac{J}{2}\mathcal{J}_{1}(\chi)\hat{\sigma}_1^{+} \hat{\sigma}_2^{+} 
+{\rm H.c.},
\end{equation}
where 
\begin{equation}
\hat{B}_\text{locked}=\mathrm{u}_1\hat{b}+\mathrm{v}_1\hat{b}^{\dagger}\label{eq_Blocked}
\end{equation}
is the phase-locked Bogoliubov squeezed operator, with
\begin{eqnarray}
\mathrm{u}_1&=&\left(\mathcal{J}_{0}(\chi)-\mathcal{J}_{1}(\chi)\right)/N_1,\\
\mathrm{v}_1&=&\left(\mathcal{J}_{2}(\chi)-\mathcal{J}_{1}(\chi)\right)/N_1,\\
N_1&=&\sqrt{\vert\left(\mathcal{J}_{0}(\chi)-\mathcal{J}_{1}(\chi)\right)^2-\left(\mathcal{J}_{2}(\chi)-\mathcal{J}_{1}(\chi)\right)^2\vert}.
\end{eqnarray}
In the above, we have assumed that the parameters $\mathrm{u}_1$ and $\mathrm{v}_1$ satisfy the Bogoliubov relation $|\mathrm{u}_1|^2 - |\mathrm{v}_1|^2 = 1$ with $\vert \mathrm{u}_1\vert>\vert \mathrm{v}_1\vert$ (see Fig. \ref{figS3}\hyperref[figS3]{(b)} in Appendix ~\ref{sm4}), and that $g_{1,\text{eff}}=\tilde{g}_{\text{eff}}N_1$ is a renormalized coupling. 
Note that setting the resonance conditions $\nu=\omega$ and $\Omega=\sum_{j=1}^{2}\Delta_j + \omega$ not only enables the presence of a squeezing operator in the Hamiltonian in Eq.~\eqref{eq:heff2_ll}, but also that the second term in Eq.~\eqref{eq:heff2_ll} corresponds to a collective spin operator that breaks the phase symmetry, thereby enabling phase locking of the squeezed amplification process.

\begin{figure}[t]
\centering
\includegraphics[width=1\linewidth]{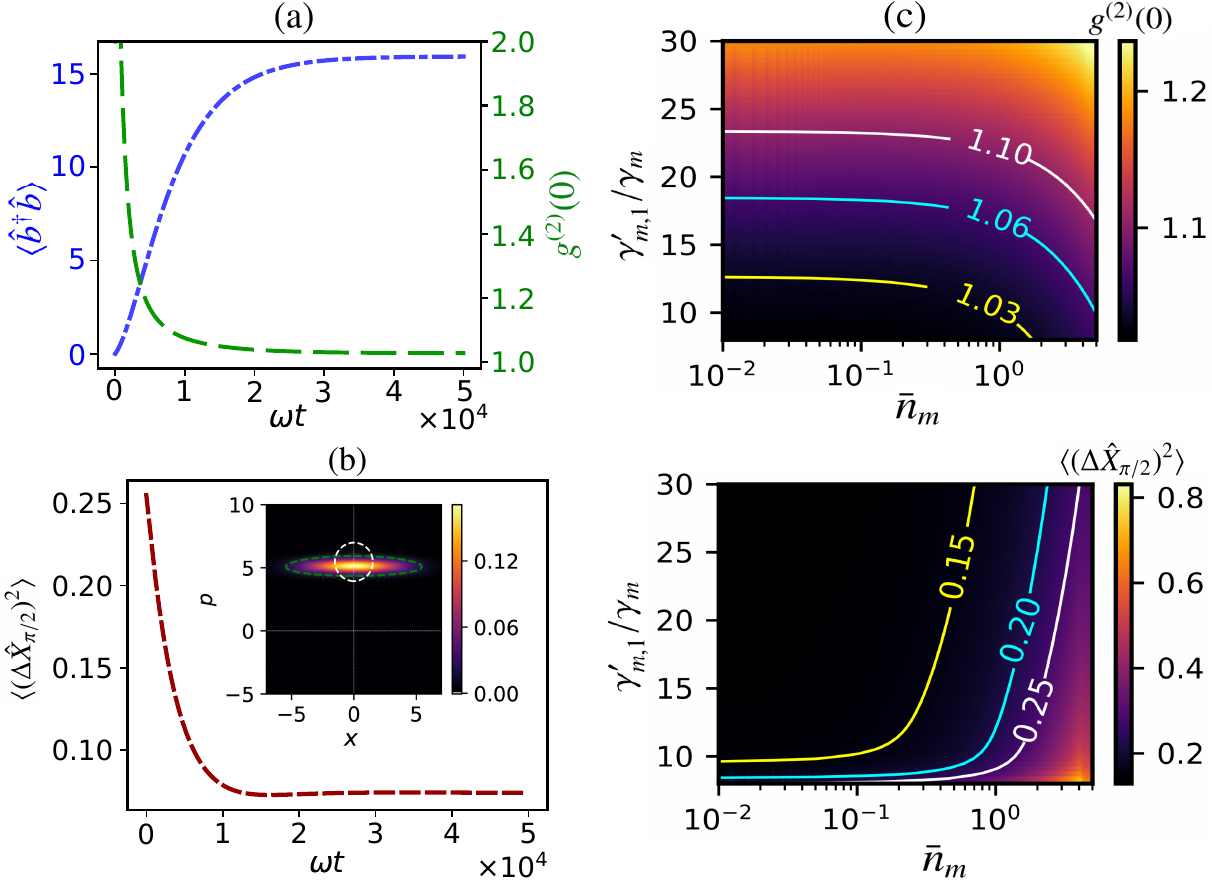}
\caption{All panels show numerical solutions of the ME in Eq.~\eqref{ME3}. 
$(a)$ Mean phonon number $\langle \hat{b}^\dagger \hat{b}\rangle$ (left y-axis) and second-order correlation function $g^{(2)}(0)$ (right y-axis) as functions of the dimensionless time $\omega t$. $(b)$ Time evolution of the quadrature fluctuation $\langle (\Delta \hat{X}_{\pi/2})^2 \rangle$, corresponding to $\theta=\pi/2$. The inset of (b) shows the Wigner function $W(x,p)$ in the steady state ($\omega t\rightarrow\infty$). The green dashed lines indicate the contour at $0.1\max[W(x,p)]$, while the white dashed lines show the corresponding contour for a coherent state with the same coherent amplitude at $\theta=\pi/2$. $(c)$ Second-order correlation function $g^{(2)}(0)$ and $(d)$ quadrature fluctuation $\langle (\Delta \hat{X}_{\pi/2})^2 \rangle$ as functions of the ratio $\gamma'_{m,1}/\gamma_m$ and the MO thermal occupation $\bar{n}_m$. The optimal regime for squeezed phonon lasing is achieved around $\gamma'_{m,1}/\gamma_m \sim 10$ and for $\bar{n}_m \lesssim 1$. Other parameters are $J=0.4\omega$, $\lambda_1=0.04\omega$, $\nu=\nu'=\omega$, $\chi = 0.8\omega$, $\Gamma_{j}=10^{-1}\omega$, $\gamma_m=2\times10^{-5}\omega$, $\gamma'_{m,1}=12\gamma_m$, $\bar{n}_{j}^{s}=10^{-5}$, and $\bar{n}_{m}=10^{-3}$.
}
\label{fig4}
\end{figure}

\subsection{Ancillary spin pair at resonance $\nu'=\omega$}
Similar to the squeezed cooling obtained in Sec. \ref{sec4}, it is possible to achieve squeezed cooling while selecting a preferred phase as discussed in the previous section. This can be realized by assuming that the auxiliary qubits are driven with the same parameters as before, except that we now set $\nu'=\omega$ and $\Omega'=\sum_{j=1}^{2}\Delta'_j - \omega$. Under these conditions, one obtains the effective Hamiltonian
\begin{eqnarray} \label{eq:heff2_l2}
\hat{H}_\text{locked}^{C}&{=}&i\tilde{g}'_{1,\text{eff}}\Bigg(\left[\mathcal{J}_{-2}(\chi){-}\mathcal{J}_{-1}(\chi)\right]\hat{b}_1^{\dagger}{+}\left[\mathcal{J}_{0}(\chi){-}\mathcal{J}_{-1}(\chi)\right]\hat{b}\Bigg)\nonumber\\
    &\times&\hat{\sigma}_1'^{+} \hat{\sigma}_2'^{+}+\frac{J'}{2}\mathcal{J}_{-1}(\chi)\hat{\sigma}_1'^{+} \hat{\sigma}_2'^{+}+\text{H.c.}.
\end{eqnarray}
Without loss of generality, to avoid the asymmetry associated with odd-order Bessel coefficients, we choose opposite drive amplitudes, $\varepsilon'=-\varepsilon$. Under this choice, we can use the symmetry relation $\mathcal{J}_{-n}(x)=(-1)^n\mathcal{J}_n(x)$. Therefore, the Hamiltonian reduces to
\begin{equation} 
\label{eq:heff2_l}
\hat{H}_\text{locked}^{C}=i g_{1,\text{eff}}'\hat{B}_\text{locked}\hat{\sigma}_1'^{+} \hat{\sigma}_2'^{+}+\frac{J'}{2}\mathcal{J}_{1}(\chi)\hat{\sigma}_1'^{+} \hat{\sigma}_2'^{+}+\text{H.c.},
\end{equation}
where $\hat{B}_\mathrm{locked}$ is given in Eq.~\eqref{eq_Blocked}, $\hat{\sigma}_1'^{+}$, $\hat{\sigma}_2'^{+}$ and $g'_{1,\text{eff}}$ correspond to operators and couplings of the auxiliary spins, respectively.

As before, we perform an adiabatic elimination of the ancilla degrees of freedom in $\hat{H}_\text{locked}^{C}$, assuming a fast decay of the ancilla spins~\cite{petruccione2002theory, Porras2012}. Therefore, the ancilla spins remain close to their ground state and act as an effective dissipative reservoir, leading to the reduced ME
\begin{eqnarray}\label{ME3}
     \frac{d\hat{\rho}}{dt}&=&-i[\hat{H}_\text{locked}^{A},\hat{\rho}] + \gamma'_{m,1}\mathcal{L}_{\hat{B}_\text{locked}}[\hat{\rho}]\nonumber\\
     &+&\gamma_m\left(1+\bar{n}_{m}\right)\mathcal{L}_{\hat{b}}[\hat{\rho}]+\gamma_m\bar{n}_{m}\mathcal{L}_{\hat{b}^\dagger}[\hat{\rho}]\nonumber\\
     &+&\sum_{j=1}^2\Gamma_{j}\left(1+\bar{n}_{j}^{s}\right)\mathcal{L}_{\hat{\sigma}_{j}^{-}}[\hat{\rho}]+\Gamma_{j}\bar{n}_{j}^{s}\mathcal{L}_{\hat{\sigma}_{j}^{+}}[\hat{\rho}],
\end{eqnarray}
where $\hat{H}_\text{locked}^{A}$ is defined in Eq. \eqref{eq:heff2_ll}. 
Here, the damping of the squeezed mode $\hat{B}_\text{locked}$ is described by the decay rate $\gamma'_{m,1}$, with $\gamma'_{m,1} \gg \gamma_m$, so that the effective dissipator $\mathcal{L}_{\hat{B}_\text{locked}}[\hat{\rho}]$ dominates over the effective phonon dissipation $\mathcal{L}_{\hat{b}}[\hat{\rho}]$.

In Fig. \ref{fig4}\hyperref[fig4]{(a)}, we show the time evolution of the mean phonon number $\langle \hat{b}^\dagger \hat{b} \rangle$ and the second-order correlation function $g^{(2)}(0)$. Starting from the thermal mechanical state with $\bar{n}_{m}=10^{-3}$, the system exhibits a rapid increase in the phonon population, eventually reaching a steady-state value. Simultaneously, $g^{(2)}(0)\to 1$, indicating that the phonon statistics approach the Poissonian limit. Note, however, that the resulting steady state is not a phase-locked coherent phonon laser, but rather a phase-locked squeezed phonon laser. Indeed, for a phase-locked displaced squeezed state $D(\zeta)S(\xi)|0\rangle$, with real displacement amplitude $\zeta$ and real squeezing parameter $\xi$, the second-order correlation function $g^{(2)}(0)$ reads~\cite{grosse2007measuring}
\begin{equation}
g^{(2)}(0)=1+\frac{\sinh^2(\xi)\left[2\zeta^2+\cosh(2\xi)-2\zeta^2\coth(\xi)\right]}{\left[\zeta^2+\sinh^2(\xi)\right]^2}.
\end{equation}
This expression shows that when the coherent displacement dominates, $\zeta^2\gg\sinh^2\xi$, one obtains $g^{(2)}(0)\to 1$---as shown in Fig.~\ref{fig4}\hyperref[fig4]{(a)}. In contrast, when the squeezed fluctuations are relevant, the state can exhibit either bunching or antibunching, depending on the relative strength of $\zeta$ and $\xi$. 

To quantify the squeezing of the generated phonon-lasing state, we evaluate the variance of the generalized quadrature operator. The quadrature squeezing is characterized by the variance
\begin{equation}
\langle (\Delta \hat{X}_{\theta})^2 \rangle = \langle \hat{X}_{\theta}^2 \rangle - \langle \hat{X}_{\theta} \rangle^2,
\label{eq:QF}
\end{equation}
where the generalized quadrature operator is defined as
\begin{equation}
\hat{X}_{\theta}=\frac{1}{2}\left(\hat{b}e^{-i\theta}+\hat{b}^{\dagger}e^{i\theta}\right).
\end{equation}
The angle $\theta$ specifies the direction in phase space along which the fluctuations are evaluated, so that for the Wigner function $W(x,p)$, one has $\hat{x}=\hat{X}_0$ and $\hat{p}=\hat{X}_{\pi/2}$. Within this convention, the phonon mode is said to be squeezed when $\langle (\Delta \hat{X}_{\theta})^2 \rangle < 1/4$, corresponding to fluctuations below the standard quantum limit.

In Fig. \ref{fig4}\hyperref[fig4]{(b)}, we present the time evolution of the quadrature variance $\langle (\Delta \hat{X}_{\theta})^2 \rangle$ for $\theta=\pi/2$. The variance is initially above the standard quantum limit but decreases over time 
entering the squeezed regime and stabilizing below the vacuum threshold, thus indicating steady-state squeezing. In the standard mechanical basis, the coherent state turns into a squeezed coherent state as shown in inset Fig. \ref{fig4}\hyperref[fig4]{(b)}, which clearly exhibits an elongated distribution along one quadrature, confirming the presence of squeezed phonon lasing. The green dashed contour highlights the squeezed profile, while the white dashed contour corresponds to a coherent state with the same amplitude, therefore emphasizing the non-classical character of the lasing steady state.


To investigate the resilience of the squeezed lasing against the detrimental effects of the thermal reservoir, we analyze in Figs.~\ref{fig4}\hyperref[fig4]{(c)-(d)} the dependence of the second-order coherence function $g^{(2)}(0)$ and the squeezing witness $\langle (\Delta \hat{X}_{\theta})^2 \rangle$ on both the mean thermal phonon number $\bar{n}_{m}$ and the engineered decay ratio $\gamma'_{m,1}/\gamma_{m}$. Remarkably, the figures reveal that the squeezed lasing regime remains robust and well-protected against thermal fluctuations up to $\bar{n}_{m} \sim 1$, provided that the ratio $\gamma'_{m,1}/\gamma_{m}$ is appropriately controlled. This result demonstrates that the squeezed lasing behavior emerging from the proposed model exhibits a strong resilience to thermal noise, suggesting its practical viability even under realistic experimental conditions where complete ground-state cooling of the mechanical mode may not be achievable.

\vspace{1cm}
\section{Conclusions}
\label{sec7}

In this work, we propose a general Floquet-engineered framework for squeezed phonon lasing that enables a continuous crossover from conventional phase-diffused phonon lasing to phase-locked squeezed phonon lasing. Our system consists of two pairs of periodically driven spins---a primary pair and an ancilla pair---coupled to a single mechanical mode. While the underlying mechanism applies generally to such spin-boson systems, we focus on a concrete implementation based on color centers embedded in a circular hexagonal boron nitride (hBN) membrane. Our steady-state analysis reveals clear signatures of phonon lasing, including a well-defined threshold, enhanced mechanical occupation, linewidth narrowing, and nontrivial second-order correlations. We show that the primary spins provide mechanical gain, whereas the ancilla spins engineer a dissipative channel for the Bogoliubov mode, thereby stabilizing steady-state squeezed phonon lasing. Moreover, the interplay between local spin driving, periodically modulated spin-spin interactions, and spin-mechanical coupling under suitable Floquet resonances gives rise to a rich dynamical regime where squeezed amplification and squeezed cooling coexist. Remarkably, the Floquet protocol intrinsically induces phase locking, suppressing phase diffusion and thereby stabilizing steady-state squeezed phonon lasing without external feedback or optical and microwave cavities. Beyond the regime explored here, this framework naturally extends to the preparation of other nonclassical mechanical states in hBN platforms and may also enable applications in quantum sensing, for example through the detection of external fields (see Appendix~\ref{appendix_metrology} for a brief discussion of AC sensing).

\vspace{-0.4cm}
\section*{ACKNOWLEDGMENTS}
\label{sec8}

H.M. acknowledges Universidad de La Frontera and partial financial support from the project ``FRO2395”, from the Ministry of Education of Chile. 
G.R. acknowledges financial support from the Provincia Autonoma di Trento (PAT).
V.M. thanks support from Khalifa University of Science and Technology through the Project ID: KU-INT-RIG-2024-8474000739 and from the National Natural Science Foundation of China Grants No. 12374482 and No. W2432005.

\onecolumngrid

\newpage

\renewcommand{\appendixname}{APPENDIX}

\appendix

\section{EFFECTIVE HAMILTONIANS FOR THE FLOQUET CONTROL OF LASERS}
\label{sect_S1}
Let us consider the Hamiltonian described in Sec. \ref{sec2a}. The Hamiltonian reads (with $\hbar=1$): 
\begin{align}\label{eq_local_driven_hamiltonian}
    \hat{H}&=\hat{H}_S+\hat{H}_J(t)+\hat{H}_{LD}(t),       
\end{align}
where $\hat{H}_S=\sum_{j=1}^{2}\frac{\Delta_j}{2} \hat{\sigma}_{j}^{z} + \omega\hat{b}^{\dagger}\hat{b}-\lambda_1\hat{\sigma}_{1}^{z}(\hat{b}^{\dagger}+\hat{b})$, $\hat{H}_J(t)=J\cos{(\Omega t)}\hat{\sigma}_{1}^{x}\hat{\sigma}_{2}^{x}$ and $\hat{H}_{LD}(t)=\varepsilon\cos{(\nu t)} \hat{\sigma}_{2}^{z}$. Here, $\varepsilon$ and $\nu$ denote the amplitude and frequency of the local driving field, respectively.

\subsection*{Derivation of the effective Hamiltonian}
To derive the effective Hamiltonian, we move to the rotating frame with respect to $\hat{H}_{LD}(t)$. Since $[\hat{H}_S,\hat{H}_{LD}(t)]=0$, and the Hamiltonian $\hat{H}_{LD}(t)$ commutes with itself at different times $t_1\ne t_2$, that is, $[\hat{H}_{LD}(t_1),\hat{H}_{LD}(t_2)]=0$, the unitary transformation reads $\mathcal{\hat{U}}(t)=e^{-\frac{i}{\hbar}\int_{0}^{t}\hat{H}_{LD}(s)ds}$. This leads us to the Hamiltonian in the rotating frame
\begin{eqnarray}
\hat{H}_{rot,1}&=&\hat{H}_S+J\cos{(\Omega t)}(\sigma_1^{+}+\sigma_1^{-})(e^{2i\mathcal{F}(t)}\sigma^{+}_2+e^{-2i\mathcal{F}(t)}\sigma^{-}_2),
\end{eqnarray}
where $\mathcal{F}(t)=(\varepsilon/\nu)\sin{\nu t}$. Now, we make use of the Jacobi-Anger expansion $e^{iz\sin{\theta}}=\sum_{m\in\mathbf{Z}}\mathcal{J}_m(z)e^{im\theta}$, where $\mathcal{J}_m(z)$ are first-order Bessel functions, and rewrite the Hamiltonian as follows
\begin{equation}
    \hat{H}_{{\rm rot},1}(t)= \hat{H}_S+J\cos{(\Omega t)}(\sigma^{+}_1+\sigma^{-}_1)\bigg(\sum_{m\in\mathbf{Z}}\mathcal{J}_m(2\varepsilon/\nu)e^{i m\nu t}\sigma^{+}_2+\sum_{m\in\mathbf{Z}}\mathcal{J}_m(2\varepsilon/\nu)e^{-i m\nu t}\sigma^{-}_2\bigg).
\end{equation}
Now, we move to the second interaction picture: $\hat{H}_{rot,2}=e^{i \hat{H}'_{0} t}\hat{H}'_{1}e^{-i \hat{H}'_{0} t}$, where 
\begin{eqnarray}
\hat{H}'_{0}&=&\sum_{j=1}^{2}\frac{\Delta_j}{2} \hat{\sigma}_{j}^{z} 
    + \omega\hat{b}^\dagger \hat{b}, \\ \hat{H}'_{1}&=&J\cos{(\Omega t)}(\sigma^{+}_1+\sigma^{-}_1)\bigg(\sum_{m\in\mathbf{Z}}\mathcal{J}_m(2\varepsilon/\nu)e^{i m\nu t}\sigma^{+}_2+\sum_{m\in\mathbf{Z}}\mathcal{J}_m(2\varepsilon/\nu)e^{-i m\nu t}\sigma^{-}_2\bigg)-\lambda_{1}\hat{\sigma}_{1}^{z}(\hat{b}^{\dagger}+\hat{b}).
\end{eqnarray}

With the above Hamiltonians and interaction picture transformation, we define $\hat{H}_{rot,2}=\hat{\mathcal{V}}_{0}+\hat{\mathcal{V}}_{1}$. By performing the interaction-picture transformation with respect to $\hat H'_0$, we additionally apply the bosonic phase rotation $\hat{U}=e^{-i(\pi/2)\hat{b}^\dagger \hat{b}}$, which transforms $\hat{b}\rightarrow i\hat{b}$ and $\hat{b}^\dagger\rightarrow -i\hat{b}^\dagger$. This yields 
\begin{align}
    \hat{\mathcal{V}}_{0}&=i\lambda_1\hat{\sigma}_{1}^{z}\left(\hat{b}^{\dagger}e^{i\omega t}-\hat{b}e^{-i\omega t}\right)\equiv\hat{\sigma}_{1}^{z}\hat{f}(t),\\
    \hat{\mathcal{V}}_{1}&=J\cos{(\Omega t)}\bigg(\sum_{m\in\mathbf{Z}}\mathcal{J}_m(2\varepsilon/\nu)e^{i m\nu t}e^{i(\Delta_1+\Delta_2)t}\sigma^{+}_1\sigma^{+}_2+\sum_{m\in\mathbf{Z}}\mathcal{J}_m(2\varepsilon/\nu)e^{-i m\nu t}e^{i(\Delta_1-\Delta_2)t}\sigma^{+}_1\sigma^{-}_2\bigg) + {\rm H.c.},
\end{align}
with $\hat{f}(t)=i\lambda_1\left(\hat{b}^{\dagger}e^{i\omega t}-\hat{b}e^{-i\omega t}\right)$. 

Next, we move to a third interaction picture, which is defined as follows:
\begin{eqnarray}\label{2ndnew0}    \hat{\mathcal{V}}'&=&\exp{\left\{i\int\hat{\mathcal{V}}_{0}dt\right\}}\hat{\mathcal{V}}_{1}\exp{\left\{-i\int\hat{\mathcal{V}}_{0}dt\right\}}\nonumber\\
&=&J\cos{(\Omega t)}\Big(\sum_{m\in\mathbf{Z}}\mathcal{J}_m(2\varepsilon/\nu)e^{i m\nu t}e^{i(\Delta_1+\Delta_2)t}e^{i\hat{\sigma}_{1}^{z}\hat{F}(t)}\hat{\sigma}_1^{+} \hat{\sigma}_2^{+}e^{-i\hat{\sigma}_{1}^{z}\hat{F}(t)}+\sum_{m\in\mathbf{Z}}\mathcal{J}_m(2\varepsilon/\nu)e^{-i m\nu t}e^{i(\Delta_1-\Delta_2)t}e^{i\hat{\sigma}_{1}^{z}\hat{F}(t)}\hat{\sigma}_1^{+} \hat{\sigma}_2^{-}e^{-i\hat{\sigma}_{1}^{z}\hat{F}(t)}\Big)+ \text{H.c.}\nonumber\\
&\equiv& \hat{\mathcal{V}}'_{I}+\hat{\mathcal{V}}'_{II} + \text{H.c.},
\end{eqnarray}
with the Hermitian operator $\hat{F}(t)\equiv\int_0^t \hat{f}(t') dt'=\frac{\lambda_1}{\omega}\left(\hat{b}^{\dagger}\varsigma+\hat{b}\varsigma^{*}\right)$, where $\varsigma=e^{i\omega t}-1$. In deriving the effective Hamiltonian, we assume that the spin-mechanical coupling strength, $\lambda_1$, is much smaller than the mechanical frequency, $\omega$. Then, expanding the exponential operator to first order in the small parameter $\lambda_1/\omega$, one has $e^{i \hat{F}(t)}\approx1+i\frac{\lambda_1}{\omega}\left(\hat{b}^{\dagger}\varsigma+\hat{b}\varsigma^{*}\right)$. 

In what follows, the explicit form of each operator $\hat{\mathcal{V}}'$ appearing in Eq.~\eqref{2ndnew0} is derived:
\begin{eqnarray*}
    \hat{\mathcal{V}}'_{I}&\equiv& J\cos{(\Omega t)}\sum_{m\in\mathbf{Z}}\mathcal{J}_m(2\varepsilon/\nu)e^{i m\nu t}e^{i(\Delta_1+\Delta_2)t}e^{i\hat{\sigma}_{1}^{z}\hat{F}(t)}\hat{\sigma}_1^{+} \hat{\sigma}_2^{+}e^{-i\hat{\sigma}_{1}^{z}\hat{F}(t)}=J\cos{(\Omega t)}\sum_{m\in\mathbf{Z}}\mathcal{J}_m(2\varepsilon/\nu)e^{i m\nu t}e^{i(\Delta_1+\Delta_2)t}\hat{\sigma}_1^{+} \hat{\sigma}_2^{+}e^{2i\hat{F}(t)}\nonumber\\
    &=& J\cos{(\Omega t)}\sum_{m\in\mathbf{Z}}\mathcal{J}_m(2\varepsilon/\nu)e^{i m\nu t}e^{i(\Delta_1+\Delta_2)t}\hat{\sigma}_1^{+} \hat{\sigma}_2^{+}\Big[1+2i\frac{\lambda_1}{\omega}\left(\hat{b}^{\dagger}\varsigma+\hat{b}\varsigma^{*}\right)\Big]\nonumber\\
    &=& \frac{J}{2}\left(e^{i\Omega t}+e^{-i\Omega t}\right)\sum_{m\in\mathbf{Z}}\mathcal{J}_m(2\varepsilon/\nu)e^{i m\nu t}e^{i(\Delta_1+\Delta_2)t}\hat{\sigma}_1^{+} \hat{\sigma}_2^{+}\nonumber\\
    &+& i\frac{J\lambda_1}{\omega}\left(e^{i\Omega t}+e^{-i\Omega t}\right)\sum_{m\in\mathbf{Z}}\mathcal{J}_m(2\varepsilon/\nu)\Big[\hat{\sigma}_1^{+}\hat{\sigma}_2^{+}\hat{b}^{\dagger}e^{i(m\nu+\Delta_1+\Delta_2+\omega)t}-\hat{\sigma}_1^{+}\hat{\sigma}_2^{+}\left(\hat{b}^{\dagger}+\hat{b}\right)e^{i(m\nu+\Delta_1+\Delta_2)t}+\hat{\sigma}_1^{+}\hat{\sigma}_2^{+}\hat{b}e^{i(m\nu+\Delta_1+\Delta_2-\omega)t}\Big],
\end{eqnarray*}

\begin{eqnarray*}
    \hat{\mathcal{V}}'_{II}&\equiv&  J\cos{(\Omega t)}\sum_{m\in\mathbf{Z}}\mathcal{J}_m(2\varepsilon/\nu)e^{-i m\nu t}e^{i(\Delta_1-\Delta_2)t}e^{i\hat{\sigma}_{1}^{z}\hat{F}(t)}\hat{\sigma}_1^{+} \hat{\sigma}_2^{-}e^{-i\hat{\sigma}_{1}^{z}\hat{F}(t)}=J\cos{(\Omega t)}\sum_{m\in\mathbf{Z}}\mathcal{J}_m(2\varepsilon/\nu)e^{-i m\nu t}e^{i(\Delta_1-\Delta_2)t}\hat{\sigma}_1^{+} \hat{\sigma}_2^{-}e^{2i\hat{F}(t)}\nonumber\\
    &=& J\cos{(\Omega t)}\sum_{m\in\mathbf{Z}}\mathcal{J}_m(2\varepsilon/\nu)e^{-i m\nu t}e^{i(\Delta_1-\Delta_2)t}\hat{\sigma}_1^{+} \hat{\sigma}_2^{-}\Big[1+2i\frac{\lambda_1}{\omega}\left(\hat{b}^{\dagger}\varsigma+\hat{b}\varsigma^{*}\right)\Big]\nonumber\\
    &=& \frac{J}{2}\left(e^{i\Omega t}+e^{-i\Omega t}\right)\sum_{m\in\mathbf{Z}}\mathcal{J}_m(2\varepsilon/\nu)e^{-i m\nu t}e^{i(\Delta_1-\Delta_2)t}\hat{\sigma}_1^{+} \hat{\sigma}_2^{-}\nonumber\\
    &+& i\frac{J\lambda_1}{\omega}\left(e^{i\Omega t}+e^{-i\Omega t}\right)\sum_{m\in\mathbf{Z}}\mathcal{J}_m(2\varepsilon/\nu)\Big[\hat{\sigma}_1^{+}\hat{\sigma}_2^{-}\hat{b}^{\dagger}e^{i(-m\nu+\Delta_1-\Delta_2+\omega)t}-\hat{\sigma}_1^{+}\hat{\sigma}_2^{-}\left(\hat{b}^{\dagger}+\hat{b}\right)e^{i(-m\nu+\Delta_1-\Delta_2) t}+\hat{\sigma}_1^{+}\hat{\sigma}_2^{-}\hat{b}e^{i(-m\nu+\Delta_1-\Delta_2-\omega)t}\Big].
\end{eqnarray*}

In the following, we apply the Floquet theory to time-periodic Hamiltonians \cite{Bukov2015AdvPhys}. The main features of the system dynamics can be captured by the one-period evolution operator $\hat{U}(T)=e^{-i\hat{H} T}$, where $\hat{H}$ is the time-independent Floquet Hamiltonian. In the high-frequency regime, $\hat{H}$ can be approximated using the Magnus expansion $\hat{H}=\sum_{l=0}^{\infty}\hat{H}^{(l)}$. The first term read as:
\begin{eqnarray*}
    \hat{H}^{(0)}&=&\frac{1}{T}\int_0^T dt \hat{\mathcal{V}}'(t)\nonumber\\    &\simeq&\frac{J}{2}\bigg(\sum_{m\in\mathbf{Z}}\mathcal{J}_m(2\varepsilon/\nu)\delta_{m\nu+\Delta_1+\Delta_2+\Omega,0}\hat{\sigma}_1^{+} \hat{\sigma}_2^{+}+\sum_{m\in\mathbf{Z}}\mathcal{J}_m(2\varepsilon/\nu)\delta_{-m\nu+\Delta_1-\Delta_2+\Omega,0}\hat{\sigma}_1^{+} \hat{\sigma}_2^{-}\bigg)\nonumber\\    &+&\frac{J}{2}\bigg(\sum_{m\in\mathbf{Z}}\mathcal{J}_m(2\varepsilon/\nu)\delta_{m\nu+\Delta_1+\Delta_2-\Omega,0}\hat{\sigma}_1^{+} \hat{\sigma}_2^{+}+\sum_{m\in\mathbf{Z}}\mathcal{J}_m(2\varepsilon/\nu)\delta_{-m\nu+\Delta_1-\Delta_2-\Omega,0}\hat{\sigma}_1^{+} \hat{\sigma}_2^{-}\bigg)\nonumber\\
    &+& i\frac{J\lambda_1}{\omega}\hat{\sigma}_1^{+}\hat{\sigma}_2^{+}\bigg(\sum_{m\in\mathbf{Z}}\mathcal{J}_m(2\varepsilon/\nu)\delta_{m\nu+\Delta_1+\Delta_2+\omega+\Omega,0}\hat{b}^{\dagger}-\sum_{m\in\mathbf{Z}}\mathcal{J}_m(2\varepsilon/\nu)\delta_{m\nu+\Delta_1+\Delta_2+\Omega,0}\left(\hat{b}^{\dagger}+\hat{b}\right)+\sum_{m\in\mathbf{Z}}\mathcal{J}_m(2\varepsilon/\nu)\delta_{m\nu+\Delta_1+\Delta_2-\omega+\Omega,0}\hat{b}\bigg)\nonumber\\
    &+& i\frac{J\lambda_1}{\omega}\hat{\sigma}_1^{+}\hat{\sigma}_2^{+}\bigg(\sum_{m\in\mathbf{Z}}\mathcal{J}_m(2\varepsilon/\nu)\delta_{m\nu+\Delta_1+\Delta_2+\omega-\Omega,0}\hat{b}^{\dagger}-\sum_{m\in\mathbf{Z}}\mathcal{J}_m(2\varepsilon/\nu)\delta_{m\nu+\Delta_1+\Delta_2-\Omega,0}\left(\hat{b}^{\dagger}+\hat{b}\right)+\sum_{m\in\mathbf{Z}}\mathcal{J}_m(2\varepsilon/\nu)\delta_{m\nu+\Delta_1+\Delta_2-\omega-\Omega,0}\hat{b}\bigg)\\
    &+& i\frac{J\lambda_1}{\omega}\hat{\sigma}_1^{+}\hat{\sigma}_2^{-}\bigg(\sum_{m\in\mathbf{Z}}\mathcal{J}_m(2\varepsilon/\nu)\delta_{-m\nu+\Delta_1-\Delta_2+\omega+\Omega,0}\hat{b}^{\dagger}-\sum_{m\in\mathbf{Z}}\mathcal{J}_m(2\varepsilon/\nu)\delta_{-m\nu+\Delta_1-\Delta_2+\Omega,0}\left(\hat{b}^{\dagger}+\hat{b}\right)+\sum_{m\in\mathbf{Z}}\mathcal{J}_m(2\varepsilon/\nu)\delta_{-m\nu+\Delta_1-\Delta_2-\omega+\Omega,0}\hat{b}\bigg)\\
    &+& i\frac{J\lambda_1}{\omega}\hat{\sigma}_1^{+}\hat{\sigma}_2^{-}\bigg(\sum_{m\in\mathbf{Z}}\mathcal{J}_m(2\varepsilon/\nu)\delta_{-m\nu+\Delta_1-\Delta_2+\omega-\Omega,0}\hat{b}^{\dagger}-\sum_{m\in\mathbf{Z}}\mathcal{J}_m(2\varepsilon/\nu)\delta_{-m\nu+\Delta_1-\Delta_2-\Omega,0}\left(\hat{b}^{\dagger}+\hat{b}\right)+\sum_{m\in\mathbf{Z}}\mathcal{J}_m(2\varepsilon/\nu)\delta_{-m\nu+\Delta_1-\Delta_2-\omega-\Omega,0}\hat{b}\bigg) + {\rm H.c.},
\end{eqnarray*}
Therefore, retaining terms up to first order, the initial Hamiltonian can be approximated as 
\begin{eqnarray}\label{H_magnus0}
    \hat{H}\equiv\hat{H}^{(0)}&=&\frac{J}{2}\bigg(\left[\mathcal{J}_{\alpha}(2\varepsilon/\nu)+\mathcal{J}_{\alpha'}(2\varepsilon/\nu)\right]\hat{\sigma}_1^{+} \hat{\sigma}_2^{+}+\left[\mathcal{J}_{\mu}(2\varepsilon/\nu)+\mathcal{J}_{\mu'}(2\varepsilon/\nu)\right]\hat{\sigma}_1^{+} \hat{\sigma}_2^{-}\bigg)\nonumber\\
&+&i\tilde{g}_{\text{eff}}\left(\tilde{\mathrm{u}}\hat{b}^{\dagger}+\tilde{\mathrm{v}}\hat{b}\right)\hat{\sigma}_1^{+} \hat{\sigma}_2^{+}+i\tilde{g}_{\text{eff}}\left(\tilde{\mathrm{u}}'\hat{b}^{\dagger}+\tilde{\mathrm{v}}'\hat{b}\right)\hat{\sigma}_1^{+} \hat{\sigma}_2^{-}+ \text{H.c.},
\end{eqnarray}
where $\tilde{g}_{\text{eff}}=J\lambda_1/\omega$ and
\begin{align}
    \tilde{\mathrm{u}}&=\left(\mathcal{J}_{\beta}(2\varepsilon/\nu)+\mathcal{J}_{\beta'}(2\varepsilon/\nu)-\mathcal{J}_{\alpha}(2\varepsilon/\nu)-\mathcal{J}_{\alpha'}(2\varepsilon/\nu)\right),\label{coef_u}\\
    \tilde{\mathrm{v}}&=\left(\mathcal{J}_{\delta}(2\varepsilon/\nu)+\mathcal{J}_{\delta'}(2\varepsilon/\nu)-\mathcal{J}_{\alpha}(2\varepsilon/\nu)-\mathcal{J}_{\alpha'}(2\varepsilon/\nu)\right),\\
    \tilde{\mathrm{u}}'&=\left(\mathcal{J}_{\eta}(2\varepsilon/\nu)+\mathcal{J}_{\eta'}(2\varepsilon/\nu)-\mathcal{J}_{\mu}(2\varepsilon/\nu)-\mathcal{J}_{\mu'}(2\varepsilon/\nu)\right),\\
    \tilde{\mathrm{v}}'&=\left(\mathcal{J}_{\kappa}(2\varepsilon/\nu)+\mathcal{J}_{\kappa'}(2\varepsilon/\nu)-\mathcal{J}_{\mu}(2\varepsilon/\nu)-\mathcal{J}_{\mu'}(2\varepsilon/\nu)\right), \label{coef_v}
\end{align}
with $\mathbb{Z}$ numbers: $\alpha=-(\Delta_1+\Delta_2+\Omega)/\nu$, $\alpha'=-(\Delta_1+\Delta_2-\Omega)/\nu$, $\mu=(\Delta_1-\Delta_2+\Omega)/\nu$, $\mu'=(\Delta_1-\Delta_2-\Omega)/\nu$, $\beta=-(\Delta_1+\Delta_2+\omega+\Omega)/\nu$, $\beta'=-(\Delta_1+\Delta_2+\omega-\Omega)/\nu$, $\delta=-(\Delta_1+\Delta_2-\omega+\Omega)/\nu$, $\delta'=-(\Delta_1+\Delta_2-\omega-\Omega)/\nu$, $\eta=(\Delta_1-\Delta_2+\omega+\Omega)/\nu$, $\eta'=(\Delta_1-\Delta_2+\omega-\Omega)/\nu$, $\kappa=(\Delta_1-\Delta_2-\omega+\Omega)/\nu$, and $\kappa'=(\Delta_1-\Delta_2-\omega-\Omega)/\nu$.

Finally, by imposing the resonance condition
$\Omega = \sum_{j=1}^{2}\Delta_j + \omega$, 
we fix the Bessel coefficients to $\beta' = 0$, $\alpha'=\omega/\nu$ and  $\delta'=2\omega/\nu$, where $\alpha',\delta' \in \mathbb{Z}$. The remaining indices of the Bessel function become significantly larger because $\Delta_1 \gg \{\Delta_2,\omega\}$. Under these conditions, we have the effective Hamiltonian
\begin{equation}\label{H_nu_apx}
    \hat{H}\left(\frac{\nu}{\omega}\right)=J_{\text{eff}}\hat{\sigma}_1^{+} \hat{\sigma}_2^{+}+i\tilde{g}_{\text{eff}}\Bigg(\left[\mathcal{J}_{\beta'}(\chi)-\mathcal{J}_{\alpha'}(\chi)\right]{b}^{\dagger}+\left[\mathcal{J}_{\delta'}(\chi)-\mathcal{J}_{\alpha'}(\chi)\right]\hat{b}\Bigg)\hat{\sigma}_1^{+} \hat{\sigma}_2^{+} + \mathcal{O}\left(\epsilon^2\right) + \text{H.c.},
\end{equation}
where $\varepsilon \equiv (\nu/2)\chi$,  $J_{\text{eff}}=J\mathcal{J}_{\alpha'}(\chi)/2$, $\epsilon=\lambda_1/\omega$, and $\tilde{g}_{\text{eff}}=J\lambda_1/\omega$. Second-order corrections $\mathcal{O}(\epsilon^2)$ are retained but not written explicitly.


\section{PARAMETERS FEASIBILITY}
\label{sm4}
\begin{figure}[b]
\centering
\includegraphics[width=0.65\linewidth]{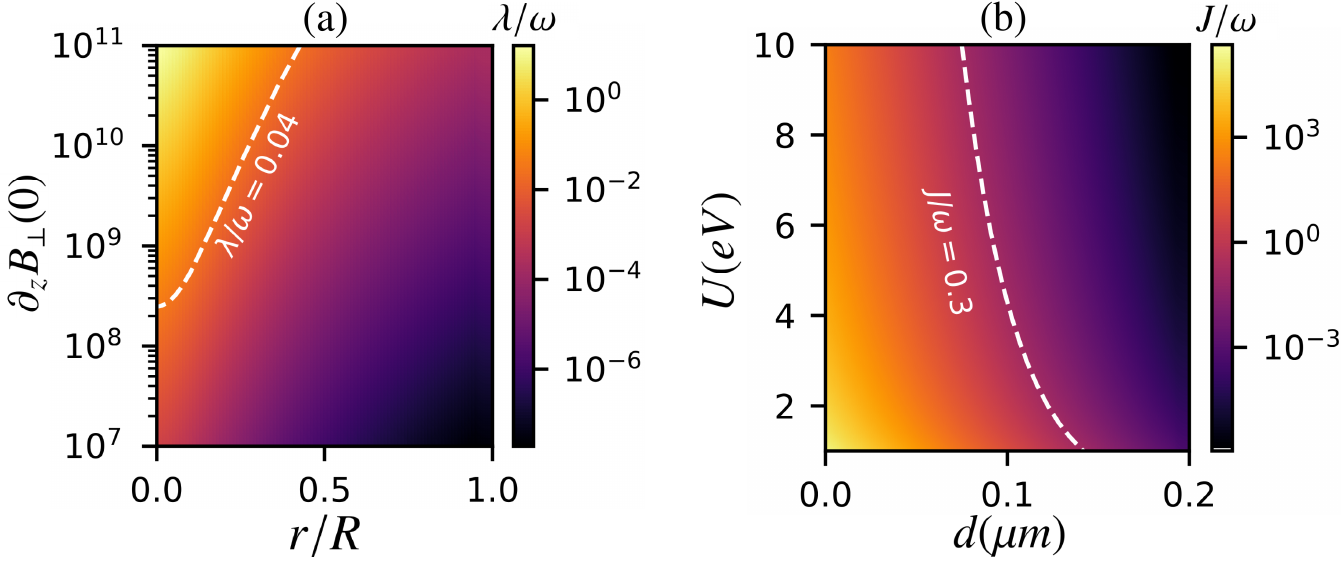}
\caption{$(a)$ Spin-motion coupling, $\lambda/\omega$, in log-scale as a function of field gradient $\partial_z B_{\perp}(0)$ and distance from the center of membrane $r$ for membrane radii $R=0.1\mu m$. $(b)$ spin-spin coupling, $J/\omega$, in log-scale as function of of the on-site repulsion $U$ and spins distance $d$. The white dashed curves correspond to the parameters used in our simulations.}
\label{fig7}
\end{figure}

In this section, we assess how the spin-motion and spin-spin coupling parameters used in the main text can be experimentally realized: 

\textit{Spin-motion coupling---} Based on Ref. \cite{PRL_hBN}, the spin-motion coupling is defined as:
\begin{equation*}
\lambda_j\equiv\lambda(r_j,\theta_j)=\mu_B g_e \partial_z B_{\perp}(r_j,\theta_j,0)z_{\circ,j}
\end{equation*}
where $\mu_B$ and $g_e$ are the Bohr magneton and the electron gyromagnetic ratio, respectively. The quantity $\partial_z B_{\perp}(r_j,\theta_j,0)$ denotes the magnetic field gradient evaluated at the position of the defect. The zero-point fluctuation amplitude of the membrane at the defect position is given by $z_{0,j}=\Psi(r_j,\theta_j)\sqrt{\hbar/(2m\omega)}$, where $\Psi(r_j,\theta_j)$ is the normalized mode profile. Here, $m=0.27(\pi R^2)\rho h$ is the effective mass of the membrane, with $R$ the membrane radius, $\rho$ the mass density, and $h$ the membrane thickness. $\Psi(r_j,\theta_j)$ is the mechanical mode profile function. For the fundamental mode, it takes the form $\Psi(r)=C_0 J_{0}(\alpha_0\frac{r}{R})$, where $J_{0}$ is the Bessel function of the first kind and $\alpha_0=2.4048$ is its first zero. Here, $C_0$ is a normalization constant. Without loss of generality, we consider a magnetic field gradient with a typical decaying spatial profile, $\partial_z B_{\perp}(r_j)=\partial_z B_{\perp}(0)/\left(1+r/d_0\right)^{7/2}$ (used in Ref. \cite{PRL_hBN}), where $d_0\sim20\mu m$ is the tip distance below the center of membrane. On the other hand, the mechanical frequency is given by:
\begin{equation*}
    \omega=2.4\sqrt{\frac{T_0}{\rho h R^2}},
\end{equation*}
where $T_0$ is the tension membrane. For hBN membranes, we consider the parameters $\rho = 2100\mathrm{kg/m^3}$, 
$h = 3.3\times10^{-10}\mathrm{m}$, $T_0 = 0.00583\mathrm{N/m}$, and $R = 0.1\mu\mathrm{m}$. These values yield a mechanical frequency $\omega \sim 350\mathrm{MHz}$, which is experimentally accessible at temperatures as low as $T = 10\mathrm{mK}$. In Fig. \ref{fig7}\hyperref[fig7]{(a)}, we observe that the spin-motion ratio $\lambda/\omega=0.04$ considered in our simulations is achievable for a magnetic field gradient above $\partial_z B_{\perp}(0)\sim10^8\mathrm{T/m}$.

\begin{figure}[h]
\centering
\includegraphics[width=0.65\linewidth]{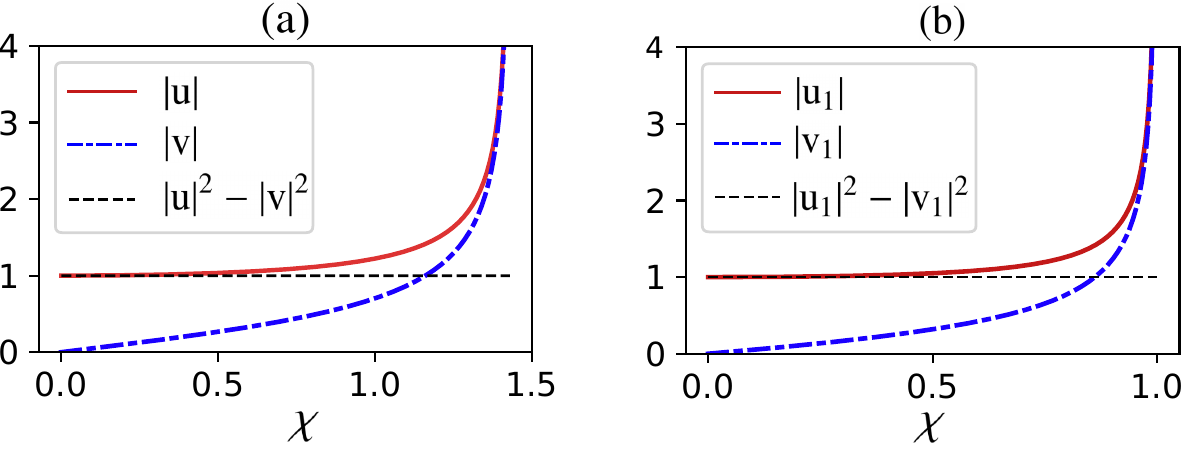}
\caption{Bogoliubov coefficients characterizing the squeezed modes: $(a)$ $\hat{B}$ and $(b)$ $\hat{B}_{\text{locked}}$. The remaining parameters are the same as those used in the main text.}
\label{figS3}
\end{figure}

\textit{Spin-spin coupling---} Based on Ref. \cite{PRB_Mod}, the spin-spin exchange coupling modulation is mediated through an effective substrate. The exchange coupling amplitude in the absence of an electric field is defined as \cite{Furuya2021}:
\begin{equation*}
J=\frac{4\vert t_1 t_2\vert^2}{\left(U+\Delta\right)}\left(\frac{1}{U}+\frac{1}{U+\Delta}\right)
\end{equation*}
where $U=U_1=U_2$ are the on-site Coulomb repulsions, and $t_{1(2)}$ is the hopping amplitude from orbital 1(2) of the spins to the substrate (orbital 3).

In Fig. \ref{fig7}\hyperref[fig7]{(b)}, we find that $J/\omega=0.3$ considered in our simulations is achievable for a spin separation of approximately $d=0.1\mu\mathrm{m}$ with $U=3\mathrm{eV}$. Here, we assume $t=t_1=t_2$, and the distance dependence follows an exponential form $t=t_0\exp{\left(-d/4\lambda_0\right)}$, where $\lambda_0=0.01\mu\mathrm{m}$ is the decay length. 
We also consider $\Delta=0.1\mathrm{eV}$.

\textit{Local spin drive---} As mentioned in the main text, an effective control of the lasing dynamics can be achieved through the frequency of the local spin drive, see Fig. \ref{fig01}\hyperref[fig01]{(b)}. In particular, for the study of squeezed lasing (Secs. \ref{sec_hBN}-\ref{sec_phase_control}), the local driving amplitude, defined as $\varepsilon=\chi\nu/2$, acts as an effective squeezing parameter \cite{petruccione2002theory}, provided that the operators $\hat{B}$ ($\hat{B}_{\text{locked}}$) satisfy the squeezed-mode condition $\vert \mathrm{u}\vert>\vert \mathrm{v}\vert\,(\vert \mathrm{u}_1\vert>\vert \mathrm{v}_1\vert)$. In Fig. \ref{figS3}, we evaluate the relation between the coefficients characterizing the Bogoliubov squeezed modes. For the case studied in Sec. \ref{sec_hBN}, namely $\nu=2\omega$, the squeezed mode operates properly for a local driving amplitude in the range $0<\varepsilon/\omega\lesssim0.75$, while for the phase-locked case studied in Sec. \ref{sec_phase_control} it operates in the range $0<\varepsilon/\omega\lesssim1$.

\begin{table}[t]
\begin{tabular}{|c|c|c|}
\hline\hline
$\hspace{1.2cm}$ Parameter $\hspace{1.2cm}$ & $\hspace{0.4cm}$ Symbol $\hspace{0.4cm}$ & $\hspace{0.7cm}$ Value $\hspace{0.7cm}$ \\ \hline
Mechanical frequency                        & $\omega/2\pi$                               & $350\mathrm{MHz}$                       \\ \hline
\multirow{2}{*}{Spin splitting}             & $\Delta_1, \Delta'_1$                    & $20\omega$                          \\ \cline{2-3} 
                                            & $\Delta_2, \Delta'_2$                    & $10\omega$                          \\ \hline
\multirow{2}{*}{Rabi frequency}             & $\Lambda_1, \Lambda'_1$                  & $0.1\omega$                         \\ \cline{2-3} 
                                            & $\Lambda_2, \Lambda'_2$                  & $0.1\omega$                         \\ \hline
\multirow{2}{*}{Spin-mechanical coupling}   & $\lambda_1, \lambda'_1$                  & $0.04\omega$                        \\ \cline{2-3} 
                                            & $\lambda_2, \lambda'_2$                  & $10^{-6}\omega$                     \\ \hline
\multirow{2}{*}{Driving frequency}          & $\Omega$                                 & $\sum_{j=1}^{2}\Delta_j+\omega$       \\ \cline{2-3} 
                                            & $\Omega'$                                & $\sum_{j=1}^{2}\Delta'_j-\omega$       \\ \hline
\multirow{2}{*}{Driving amplitude}          & $J$                                      & $\gg\Gamma_j$                           \\ \cline{2-3} 
                                            & $J'$                                     & $\ll\Gamma'_j$                          \\ \hline
Local drive frequency                     & $\nu, \nu'$                              & $1$--$10\omega$                        \\ \hline
Local drive amplitude                     & $\varepsilon, \varepsilon'$              & $\sim4\omega$                       \\ \hline
\multirow{2}{*}{Mechanical damping}          & $\gamma_m$                               & $\sim10^{-4}\omega$                     \\ \cline{2-3} 
                                            & $\gamma'_m$, $\gamma'_{m,1}$                              & $\sim10^{-3}\omega$                     \\ \hline
Spin decay                                 & $\Gamma_j, \Gamma'_j$                    & $\sim10^{-2}\omega$                     \\ \hline\hline
\end{tabular}
\caption{Feasible experimental parameters, in our work, for squeezed phonon laser in a circular hBN membrane with radius $R=0.1\mu\mathrm{m}$, operating in the MHz regime \cite{PRL_hBN}. Unprimed (primed) parameters correspond to primary (ancilla) spins.}
\label{table1}
\end{table}



\section{SQUEEZED PHONON LASER WITH COLOR CENTERS IN HEXAGONAL BORON NITRIDE (hBN) MEMBRANE}
\label{sec_hBN_apx}
In this section, we examine the Hamiltonian introduced in Ref.~\cite{PRL_hBN}, which includes an additional term accounting for a static magnetic field applied parallel to the plane of the membrane. Our goal is to elucidate the mechanism by which squeezed phonon amplification is achieved, mediated by spin defects embedded in a hBN membrane. To this end, we analyze the distinct roles played by the two primary spins $S_1$ and $S_2$, whose configuration is illustrated in Fig.~\ref{fighBN}. Specifically, $S_1$ couples dispersively to the mechanical mode, while $S_2$ is subject to an external coherent drive, together enabling the amplification of squeezed phononic states. The full Hamiltonian of the system reads (with $\hbar = 1$):
\begin{align}\label{full_hBN}
    \hat{H}_{\text{hBN}}&=\hat{H}_z+\hat{H}_x+\hat{H}_{LD}(t),       
\end{align}
where $\hat{H}_z=\sum_{j=1}^{2}\frac{\Delta_j}{2} \hat{\sigma}_{j}^{z} + \omega\hat{b}^{\dagger}\hat{b}-\lambda_1\hat{\sigma}_{1}^{z}(\hat{b}^{\dagger}+\hat{b})$, $\hat{H}_x=\sum_{j=1}^{2}\frac{\varLambda_j}{2} \hat{\sigma}_{j}^{x}+J\cos{(\Omega t)}\hat{\sigma}_{1}^{x}\hat{\sigma}_{2}^{x}$ and $\hat{H}_{LD}(t)=\varepsilon\cos{(\nu t)} \hat{\sigma}_{2}^{z}$. Here, $\varepsilon$ and $\nu$ denote the amplitude and frequency of the local driving field, respectively.

\subsection*{Derivation of the effective Hamiltonian}
Following the same procedure outlined in the previous section, we derive the effective Hamiltonian by transforming to the rotating frame defined by $\hat{H}_{\mathrm{LD}}(t)$. Since $[\hat{H}_z,\hat{H}_{LD}(t)]=0$, and the Hamiltonian $\hat{H}_{LD}(t)$ commutes with itself at different times $t_1\ne t_2$, that is, $[\hat{H}_{LD}(t_1),\hat{H}_{LD}(t_2)]=0$, the unitary transformation reads $\mathcal{\hat{U}}(t)=e^{-\frac{i}{\hbar}\int_{0}^{t}\hat{H}_{LD}(s)ds}$. This leads us to the Hamiltonian in the rotating frame
\begin{eqnarray}
\hat{H}_{rot,1}&=&\hat{H}_z+\frac{\varLambda_1}{2}(\sigma_1^{+}+\sigma_1^{-})+\frac{\varLambda_2}{2}(\sigma_2^{+}e^{2i\mathcal{F}(t)}+\sigma_2^{-}e^{-2i\mathcal{F}(t)})+J\cos{(\Omega t)}(\sigma_1^{+}+\sigma_1^{-})(e^{2i\mathcal{F}(t)}\sigma^{+}_2+e^{-2i\mathcal{F}(t)}\sigma^{-}_2),
\end{eqnarray}
where $\mathcal{F}(t)=(\varepsilon/\nu)\sin{\nu t}$. Now, we make use of the Jacobi-Anger expansion $e^{iz\sin{\theta}}=\sum_{m\in\mathbf{Z}}\mathcal{J}_m(z)e^{im\theta}$, where $\mathcal{J}_m(z)$ are first-order Bessel functions, and rewrite the Hamiltonian as follows
\begin{eqnarray}
    \hat{H}_{{\rm rot},1}(t)&=& \hat{H}_z+\frac{\varLambda_1}{2}(\sigma_1^{+}+\sigma_1^{-})+\frac{\varLambda_2}{2}\bigg(\sum_{m\in\mathbf{Z}}\mathcal{J}_m(2\varepsilon/\nu)e^{i m\nu t}\sigma^{+}_2+\sum_{m\in\mathbf{Z}}\mathcal{J}_m(2\varepsilon/\nu)e^{-i m\nu t}\sigma^{-}_2\bigg)\nonumber\\&+&J\cos{(\Omega t)}(\sigma^{+}_1+\sigma^{-}_1)\bigg(\sum_{m\in\mathbf{Z}}\mathcal{J}_m(2\varepsilon/\nu)e^{i m\nu t}\sigma^{+}_2+\sum_{m\in\mathbf{Z}}\mathcal{J}_m(2\varepsilon/\nu)e^{-i m\nu t}\sigma^{-}_2\bigg).
\end{eqnarray}
Now, we move to the second interaction picture: $\hat{H}_{rot,2}=e^{i \hat{H}'_{0} t}\hat{H}'_{1}e^{-i \hat{H}'_{0} t}$, where 
\begin{eqnarray}
\hat{H}'_{0}&=&\sum_{j=1}^{2}\frac{\Delta_j}{2} \hat{\sigma}_{j}^{z} 
    + \omega\hat{b}^\dagger \hat{b}, \\ \hat{H}'_{1}&=&\frac{\varLambda_1}{2}(\sigma_1^{+}+\sigma_1^{-})+\frac{\varLambda_2}{2}\bigg(\sum_{m\in\mathbf{Z}}\mathcal{J}_m(2\varepsilon/\nu)e^{i m\nu t}\sigma^{+}_2+\sum_{m\in\mathbf{Z}}\mathcal{J}_m(2\varepsilon/\nu)e^{-i m\nu t}\sigma^{-}_2\bigg)\nonumber\\&+&J\cos{(\Omega t)}(\sigma^{+}_1+\sigma^{-}_1)\bigg(\sum_{m\in\mathbf{Z}}\mathcal{J}_m(2\varepsilon/\nu)e^{i m\nu t}\sigma^{+}_2+\sum_{m\in\mathbf{Z}}\mathcal{J}_m(2\varepsilon/\nu)e^{-i m\nu t}\sigma^{-}_2\bigg)-\lambda_{1}\hat{\sigma}_{1}^{z}(\hat{b}^{\dagger}+\hat{b}).
\end{eqnarray}

With the above Hamiltonians and interaction picture transformation, one gets: $\hat{H}_{rot,2}=\hat{\mathcal{V}}_{0}+\hat{\mathcal{V}}_{1}$, where
\begin{align}
    \hat{\mathcal{V}}_{0}&=i\lambda_1\hat{\sigma}_{1}^{z}\left(\hat{b}^{\dagger}e^{i\omega t}-\hat{b}e^{-i\omega t}\right)\equiv\hat{\sigma}_{1}^{z}\hat{f}(t),\\
    \hat{\mathcal{V}}_{1}&=\frac{\varLambda_1}{2}\bigg(\sigma_1^{+}e^{i\Delta_1t}+\sigma_1^{-}e^{-i\Delta_1t}\bigg)+\frac{\varLambda_2}{2}\bigg(\sum_{m\in\mathbf{Z}}\mathcal{J}_m(2\varepsilon/\nu)e^{i m\nu t}e^{i\Delta_2t}\sigma^{+}_2+\sum_{m\in\mathbf{Z}}\mathcal{J}_m(2\varepsilon/\nu)e^{-i m\nu t}e^{-i\Delta_2t}\sigma^{-}_2\bigg)\nonumber\\&+J\cos{(\Omega t)}\bigg(\sum_{m\in\mathbf{Z}}\mathcal{J}_m(2\varepsilon/\nu)e^{i m\nu t}e^{i(\Delta_1+\Delta_2)t}\sigma^{+}_1\sigma^{+}_2+\sum_{m\in\mathbf{Z}}\mathcal{J}_m(2\varepsilon/\nu)e^{-i m\nu t}e^{i(\Delta_1-\Delta_2)t}\sigma^{+}_1\sigma^{-}_2 + {\rm H.c.}\bigg),
\end{align}
with $\hat{f}(t)=i\lambda_1\left(\hat{b}^{\dagger}e^{i\omega t}-\hat{b}e^{-i\omega t}\right)$.

Next, we move to a third interaction picture, which is defined as follows:
\begin{eqnarray}\label{2ndnew1}    \hat{\mathcal{V}}'&=&\exp{\left\{i\int\hat{\mathcal{V}}_{0}dt\right\}}\hat{\mathcal{V}}_{1}\exp{\left\{-i\int\hat{\mathcal{V}}_{0}dt\right\}}\nonumber\\
&=&\frac{\varLambda_1}{2}\sigma_1^{+}e^{i\Delta_1t}e^{i\hat{\sigma}_{1}^{z}\hat{F}(t)}\hat{\sigma}_1^{+}e^{-i\hat{\sigma}_{1}^{z}\hat{F}(t)}+\frac{\varLambda_2}{2}\sum_{m\in\mathbf{Z}}\mathcal{J}_m(2\varepsilon/\nu)e^{i m\nu t}e^{i\Delta_2t}\hat{\sigma}_2^{+}\nonumber\\&+&J\cos{(\Omega t)}\Big(\sum_{m\in\mathbf{Z}}\mathcal{J}_m(2\varepsilon/\nu)e^{i m\nu t}e^{i(\Delta_1+\Delta_2)t}e^{i\hat{\sigma}_{1}^{z}\hat{F}(t)}\hat{\sigma}_1^{+} \hat{\sigma}_2^{+}e^{-i\hat{\sigma}_{1}^{z}\hat{F}(t)}+\sum_{m\in\mathbf{Z}}\mathcal{J}_m(2\varepsilon/\nu)e^{-i m\nu t}e^{i(\Delta_1-\Delta_2)t}e^{i\hat{\sigma}_{1}^{z}\hat{F}(t)}\hat{\sigma}_1^{+} \hat{\sigma}_2^{-}e^{-i\hat{\sigma}_{1}^{z}\hat{F}(t)}\Big)+ \text{H.c.} \nonumber\\
&\equiv& \hat{\mathcal{V}}'_{I}+\hat{\mathcal{V}}'_{II}+\hat{\mathcal{V}}'_{III} + \text{H.c.},
\end{eqnarray}
with the Hermitian operator $\hat{F}(t)\equiv\int_0^t \hat{f}(t') dt'=\frac{\lambda_1}{\omega}\left(\hat{b}^{\dagger}\varsigma+\hat{b}\varsigma^{*}\right)$, where $\varsigma=e^{i\omega t}-1$. In deriving the effective Hamiltonian, we assume that the spin-mechanical coupling $\lambda_1$ is much weaker than the mechanical frequency $\omega$, enabling the following analytical simplification: $e^{i \hat{F}(t)}\approx1+i\frac{\lambda_1}{\omega}\left(\hat{b}^{\dagger}\varsigma+\hat{b}\varsigma^{*}\right)$. 

In what follows, the explicit form of each operator $\hat{\mathcal{V}}'$ appearing in Eq.~\eqref{2ndnew1} is derived:
\begin{eqnarray}
    \hat{\mathcal{V}}'_{I}&\equiv& \frac{\varLambda_1}{2}\sigma_1^{+}e^{i\Delta_1t}e^{i\hat{\sigma}_{1}^{z}\hat{F}(t)}\hat{\sigma}_1^{+}e^{-i\hat{\sigma}_{1}^{z}\hat{F}(t)}+\frac{\varLambda_2}{2}\sum_{m\in\mathbf{Z}}\mathcal{J}_m(2\varepsilon/\nu)e^{i m\nu t}e^{i\Delta_2t}\hat{\sigma}_2^{+}=\frac{\varLambda_2}{2}\sigma_2^{+}e^{i\Delta_2 t}+\frac{\varLambda_1}{2}\sum_{m\in\mathbf{Z}}\mathcal{J}_m(2\varepsilon/\nu)e^{i m\nu t}e^{i\Delta_1t}\hat{\sigma}_1^{+}e^{2i\hat{F}(t)}\nonumber\\
    &=&\frac{\varLambda_2}{2}\sum_{m\in\mathbf{Z}}\mathcal{J}_m(2\varepsilon/\nu)e^{i m\nu t}e^{i\Delta_2t}\hat{\sigma}_2^{+}+\frac{\varLambda_1}{2}e^{i\Delta_1t}\hat{\sigma}_1^{+}\Big[1+2i\frac{\lambda_1}{\omega}\left(\hat{b}^{\dagger}\varsigma+\hat{b}\varsigma^{*}\right)\Big]\nonumber\\
    &=& \frac{\varLambda_2}{2}\sum_{m\in\mathbf{Z}}\mathcal{J}_m(2\varepsilon/\nu)e^{i m\nu t}e^{i\Delta_2t}\hat{\sigma}_2^{+}+\frac{\varLambda_1}{2}e^{i\Delta_1t} \hat{\sigma}_1^{+}\nonumber\\
    &+& i\frac{\varLambda_1\lambda_1}{\omega}\Big[\hat{\sigma}_1^{+}\hat{b}^{\dagger}e^{i(m\nu+\Delta_1+\omega)t}-\hat{\sigma}_1^{+}\left(\hat{b}^{\dagger}+\hat{b}\right)e^{i(m\nu+\Delta_1)t}+\hat{\sigma}_1^{+}\hat{b}e^{i(m\nu+\Delta_1-\omega)t}\Big],
\end{eqnarray}

\begin{eqnarray}
    \hat{\mathcal{V}}'_{II}&\equiv& J\cos{(\Omega t)}\sum_{m\in\mathbf{Z}}\mathcal{J}_m(2\varepsilon/\nu)e^{i m\nu t}e^{i(\Delta_1+\Delta_2)t}e^{i\hat{\sigma}_{1}^{z}\hat{F}(t)}\hat{\sigma}_1^{+} \hat{\sigma}_2^{+}e^{-i\hat{\sigma}_{1}^{z}\hat{F}(t)}=J\cos{(\Omega t)}\sum_{m\in\mathbf{Z}}\mathcal{J}_m(2\varepsilon/\nu)e^{i m\nu t}e^{i(\Delta_1+\Delta_2)t}\hat{\sigma}_1^{+} \hat{\sigma}_2^{+}e^{2i\hat{F}(t)}\nonumber\\
    &=& J\cos{(\Omega t)}\sum_{m\in\mathbf{Z}}\mathcal{J}_m(2\varepsilon/\nu)e^{i m\nu t}e^{i(\Delta_1+\Delta_2)t}\hat{\sigma}_1^{+} \hat{\sigma}_2^{+}\Big[1+2i\frac{\lambda_1}{\omega}\left(\hat{b}^{\dagger}\varsigma+\hat{b}\varsigma^{*}\right)\Big]\nonumber\\
    &=& \frac{J}{2}\left(e^{i\Omega t}+e^{-i\Omega t}\right)\sum_{m\in\mathbf{Z}}\mathcal{J}_m(2\varepsilon/\nu)e^{i m\nu t}e^{i(\Delta_1+\Delta_2)t}\hat{\sigma}_1^{+} \hat{\sigma}_2^{+}\nonumber\\
    &+& i\frac{J\lambda_1}{\omega}\left(e^{i\Omega t}+e^{-i\Omega t}\right)\sum_{m\in\mathbf{Z}}\mathcal{J}_m(2\varepsilon/\nu)\Big[\hat{\sigma}_1^{+}\hat{\sigma}_2^{+}\hat{b}^{\dagger}e^{i(m\nu+\Delta_1+\Delta_2+\omega)t}\nonumber\\
    &-&\hat{\sigma}_1^{+}\hat{\sigma}_2^{+}\left(\hat{b}^{\dagger}+\hat{b}\right)e^{i(m\nu+\Delta_1+\Delta_2)t}+\hat{\sigma}_1^{+}\hat{\sigma}_2^{+}\hat{b}e^{i(m\nu+\Delta_1+\Delta_2-\omega)t}\Big],
\end{eqnarray}
\begin{eqnarray}
    \hat{\mathcal{V}}'_{III}&\equiv&  J\cos{(\Omega t)}\sum_{m\in\mathbf{Z}}\mathcal{J}_m(2\varepsilon/\nu)e^{-i m\nu t}e^{i(\Delta_1-\Delta_2)t}e^{i\hat{\sigma}_{2}^{z}\hat{F}(t)}\hat{\sigma}_1^{+} \hat{\sigma}_2^{-}e^{-i\hat{\sigma}_{2}^{z}\hat{F}(t)}=J\cos{(\Omega t)}\sum_{m\in\mathbf{Z}}\mathcal{J}_m(2\varepsilon/\nu)e^{-i m\nu t}e^{i(\Delta_1-\Delta_2)t}\hat{\sigma}_1^{+} \hat{\sigma}_2^{-}e^{2i\hat{F}(t)}\nonumber\\
    &=& J\cos{(\Omega t)}\sum_{m\in\mathbf{Z}}\mathcal{J}_m(2\varepsilon/\nu)e^{-i m\nu t}e^{i(\Delta_1-\Delta_2)t}\hat{\sigma}_1^{+} \hat{\sigma}_2^{-}\Big[1+2i\frac{\lambda_1}{\omega}\left(\hat{b}^{\dagger}\varsigma+\hat{b}\varsigma^{*}\right)\Big]\nonumber\\
    &=& \frac{J}{2}\left(e^{i\Omega t}+e^{-i\Omega t}\right)\sum_{m\in\mathbf{Z}}\mathcal{J}_m(2\varepsilon/\nu)e^{-i m\nu t}e^{i(\Delta_1-\Delta_2)t}\hat{\sigma}_1^{+} \hat{\sigma}_2^{-}\nonumber\\
    &+& i\frac{J\lambda_1}{\omega}\left(e^{i\Omega t}+e^{-i\Omega t}\right)\sum_{m\in\mathbf{Z}}\mathcal{J}_m(2\varepsilon/\nu)\Big[\hat{\sigma}_1^{+}\hat{\sigma}_2^{-}\hat{b}^{\dagger}e^{i(-m\nu+\Delta_1-\Delta_2+\omega)t}\nonumber\\
    &-&\hat{\sigma}_1^{+}\hat{\sigma}_2^{-}\left(\hat{b}^{\dagger}+\hat{b}\right)e^{i(-m\nu+\Delta_1-\Delta_2) t}+\hat{\sigma}_1^{+}\hat{\sigma}_2^{-}\hat{b}e^{i(-m\nu+\Delta_1-\Delta_2-\omega)t}\Big].
\end{eqnarray}

In the following, we apply the Floquet theory to time-periodic Hamiltonians \cite{Bukov2015AdvPhys}. The main features of the system dynamics can be captured by the one-period evolution operator $\hat{U}(T)=e^{-i\hat{H} T}$, where $\hat{H}$ is the time-independent Floquet Hamiltonian. In the high-frequency regime, $\hat{H}$ can be approximated using the Magnus expansion $\hat{H}=\sum_{l=0}^{\infty}\hat{H}^{(l)}$. The first term read as:
\begin{eqnarray*}
    \hat{H}^{(0)}&=&\frac{1}{T}\int_0^T dt \hat{\mathcal{V}}'(t)\nonumber\\    &\simeq&\frac{\varLambda_2}{2}\sum_{m\in\mathbf{Z}}\mathcal{J}_m(2\varepsilon/\nu)\hat{\sigma}_2^{+}\delta_{m\nu+\Delta_2,0}+\frac{\varLambda_1}{2}\hat{\sigma}_1^{+}\delta_{\Delta_1,0}\nonumber\\
    &+& i\frac{\varLambda_1\lambda_1}{\omega}\Big[\hat{\sigma}_1^{+}\hat{b}^{\dagger}\delta_{\Delta_1+\omega,0}-\hat{\sigma}_1^{+}\left(\hat{b}^{\dagger}+\hat{b}\right)\delta_{\Delta_1,0}+\hat{\sigma}_1^{+}\hat{b}\delta_{\Delta_1-\omega,0}\Big]\nonumber\\ &+&\frac{J}{2}\bigg(\sum_{m\in\mathbf{Z}}\mathcal{J}_m(2\varepsilon/\nu)\delta_{m\nu+\Delta_1+\Delta_2+\Omega,0}\hat{\sigma}_1^{+} \hat{\sigma}_2^{+}+\sum_{m\in\mathbf{Z}}\mathcal{J}_m(2\varepsilon/\nu)\delta_{-m\nu+\Delta_1-\Delta_2+\Omega,0}\hat{\sigma}_1^{+} \hat{\sigma}_2^{-}\bigg)\nonumber\\    &+&\frac{J}{2}\bigg(\sum_{m\in\mathbf{Z}}\mathcal{J}_m(2\varepsilon/\nu)\delta_{m\nu+\Delta_1+\Delta_2-\Omega,0}\hat{\sigma}_1^{+} \hat{\sigma}_2^{+}+\sum_{m\in\mathbf{Z}}\mathcal{J}_m(2\varepsilon/\nu)\delta_{-m\nu+\Delta_1-\Delta_2-\Omega,0}\hat{\sigma}_1^{+} \hat{\sigma}_2^{-}\bigg)\nonumber\\
    &+& i\frac{J\lambda_1}{\omega}\hat{\sigma}_1^{+}\hat{\sigma}_2^{+}\bigg(\sum_{m\in\mathbf{Z}}\mathcal{J}_m(2\varepsilon/\nu)\delta_{m\nu+\Delta_1+\Delta_2+\omega+\Omega,0}\hat{b}^{\dagger}-\sum_{m\in\mathbf{Z}}\mathcal{J}_m(2\varepsilon/\nu)\delta_{m\nu+\Delta_1+\Delta_2+\Omega,0}\left(\hat{b}^{\dagger}+\hat{b}\right)+\sum_{m\in\mathbf{Z}}\mathcal{J}_m(2\varepsilon/\nu)\delta_{m\nu+\Delta_1+\Delta_2-\omega+\Omega,0}\hat{b}\bigg)\nonumber\\
    &+& i\frac{J\lambda_1}{\omega}\hat{\sigma}_1^{+}\hat{\sigma}_2^{+}\bigg(\sum_{m\in\mathbf{Z}}\mathcal{J}_m(2\varepsilon/\nu)\delta_{m\nu+\Delta_1+\Delta_2+\omega-\Omega,0}\hat{b}^{\dagger}-\sum_{m\in\mathbf{Z}}\mathcal{J}_m(2\varepsilon/\nu)\delta_{m\nu+\Delta_1+\Delta_2-\Omega,0}\left(\hat{b}^{\dagger}+\hat{b}\right)+\sum_{m\in\mathbf{Z}}\mathcal{J}_m(2\varepsilon/\nu)\delta_{m\nu+\Delta_1+\Delta_2-\omega-\Omega,0}\hat{b}\bigg)\\
    &+& i\frac{J\lambda_1}{\omega}\hat{\sigma}_1^{+}\hat{\sigma}_2^{-}\bigg(\sum_{m\in\mathbf{Z}}\mathcal{J}_m(2\varepsilon/\nu)\delta_{-m\nu+\Delta_1-\Delta_2+\omega+\Omega,0}\hat{b}^{\dagger}-\sum_{m\in\mathbf{Z}}\mathcal{J}_m(2\varepsilon/\nu)\delta_{-m\nu+\Delta_1-\Delta_2+\Omega,0}\left(\hat{b}^{\dagger}+\hat{b}\right)+\sum_{m\in\mathbf{Z}}\mathcal{J}_m(2\varepsilon/\nu)\delta_{-m\nu+\Delta_1-\Delta_2-\omega+\Omega,0}\hat{b}\bigg)\\
    &+& i\frac{J\lambda_1}{\omega}\hat{\sigma}_1^{+}\hat{\sigma}_2^{-}\bigg(\sum_{m\in\mathbf{Z}}\mathcal{J}_m(2\varepsilon/\nu)\delta_{-m\nu+\Delta_1-\Delta_2+\omega-\Omega,0}\hat{b}^{\dagger}-\sum_{m\in\mathbf{Z}}\mathcal{J}_m(2\varepsilon/\nu)\delta_{-m\nu+\Delta_1-\Delta_2-\Omega,0}\left(\hat{b}^{\dagger}+\hat{b}\right)+\sum_{m\in\mathbf{Z}}\mathcal{J}_m(2\varepsilon/\nu)\delta_{-m\nu+\Delta_1-\Delta_2-\omega-\Omega,0}\hat{b}\bigg) + \text{H.c.},
\end{eqnarray*}
Therefore, retaining terms up to first order, the initial Hamiltonian can be approximated as 
\begin{eqnarray}\label{H_magnus2}
    \hat{H}\equiv\hat{H}^{(0)}&=&\frac{\varLambda_2}{2}\sigma_2^{+}\mathcal{J}_{a}(2\varepsilon/\nu)+\frac{\varLambda_1}{2}\hat{\sigma}_1^{+}\delta_{\Delta_1,0}+ i\frac{\varLambda_1\lambda_1}{\omega}\Big[\hat{\sigma}_1^{+}\hat{b}^{\dagger}\delta_{\Delta_1+\omega,0}-\hat{\sigma}_1^{+}\left(\hat{b}^{\dagger}+\hat{b}\right)\delta_{\Delta_1,0}+\hat{\sigma}_1^{+}\hat{b}\delta_{\Delta_1-\omega,0}\Big]\nonumber\\&+&\frac{J}{2}\bigg(\left[\mathcal{J}_{\alpha}(2\varepsilon/\nu)+\mathcal{J}_{\alpha'}(2\varepsilon/\nu)\right]\hat{\sigma}_1^{+} \hat{\sigma}_2^{+}+\left[\mathcal{J}_{\mu}(2\varepsilon/\nu)+\mathcal{J}_{\mu'}(2\varepsilon/\nu)\right]\hat{\sigma}_1^{+} \hat{\sigma}_2^{-}\bigg)\nonumber\\
    &+&i\tilde{g}_{\text{eff}}\left(\tilde{\mathrm{u}}\hat{b}^{\dagger}+\tilde{\mathrm{v}}\hat{b}\right)\hat{\sigma}_1^{+} \hat{\sigma}_2^{+}+i\tilde{g}_{\text{eff}}\left(\tilde{\mathrm{u}}'\hat{b}^{\dagger}+\tilde{\mathrm{v}}'\hat{b}\right)\hat{\sigma}_1^{+} \hat{\sigma}_2^{-}+ \text{H.c.},
\end{eqnarray}


\begin{figure}[b]
\centering
\includegraphics[width=0.75\linewidth]{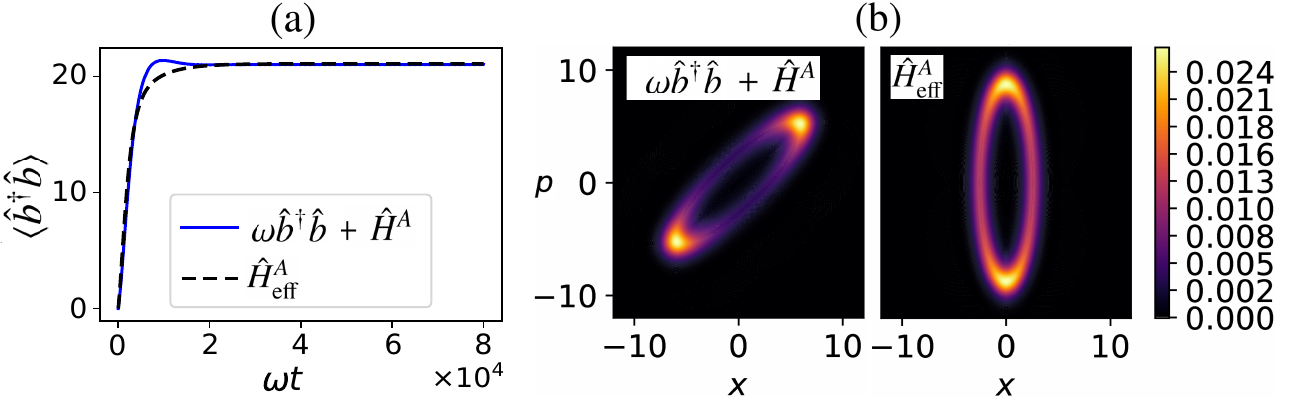}
\caption{(a) The mean phonon number ${\langle \hat{b}^{\dagger}\hat{b} \rangle}$ as function of dimensionless time, at the resonance conditions $\Omega=\sum_{j=1}^{2}\Delta_j+\omega$ and $\nu=2\omega$, as obtained by the numerical solution of the ME with the time-dependent Hamiltonian $\hat{H}_{\text{hBN}}$ defined in Eq. \eqref{full_hBN} (solid blue curve) and using the effective Hamiltonian $\hat{H}_\text{eff}^A\equiv\hat{H}_2(\nu/\omega=2)$ (dashed black line). $(b)$ Wigner function $W(x,p)$ at $t\rightarrow\infty$. Parameters: $\Delta_1=20\omega$, $\Delta_2=10\omega$, $\Lambda_1=\Lambda_2=0.1\omega$,
$\lambda_1=0.04\omega$, $\lambda_2=10^{-6}\omega$, $\nu=2\omega$, $\Omega=\Delta_1+\Delta_2+\omega$, $J=0.3\omega$, $\chi=1$, $\Gamma_{j}=0.03\omega$, $\gamma_m=10^{-3}\omega$, $\bar{n}_{j}^{s}=10^{-5}$, $\bar{n}_{m}=10^{-3}$.}
\label{fig2s}
\end{figure}

where $\tilde{g}_{\mathrm{eff}} = J\lambda_1/\omega$, $a = -\Delta_2/\nu$, and the definitions for $\tilde{\mathrm{u}}$, $\tilde{\mathrm{v}}$, $\tilde{\mathrm{u}}'$, and $\tilde{\mathrm{v}}'$ are given in Eqs. \eqref{coef_u}-\eqref{coef_v}. In what follows, we invoke the resonance conditions established in Sec.~\ref{sect_S1}, supplemented by the off-resonance assumptions $\Delta_1 \gg \Lambda_1$ and $\Delta_2 \gg \nu$. Within this regime, the Hamiltonian in Eq.~\eqref{H_magnus2} reduces to the simplified form given in Eq.~\eqref{H_nu_apx}. This result demonstrates that the hBN membrane coupled to the spin degrees of freedom is fully captured within the general theoretical framework introduced in Sec.~\ref{sect_S1}, with the initial Hamiltonian as defined in Eq.~\eqref{eq_local_driven_hamiltonian}, thereby validating the applicability of that framework to this specific physical platform.

Figure \ref{fig2s}\hyperref[fig2s]{(a)} shows the dynamics of the average phonon number under these resonance conditions for the time-dependent Hamiltonian, calculated for typical parameters of the hBN membrane (Table \ref{table1}), along with the prediction of the effective model where $\hat{H}_{\mathrm{eff}}^{A}\equiv\hat{H}_2(\nu/\omega=2)$ defined in Eq. \eqref{eq:heff1_l}.
The two solutions agree closely, confirming that phonon amplification saturates toward a steady-state value. As discussed at the end of Sect.~\ref{subsect:Squeezing-amplification}, the Wigner representation reveals that the system evolves into a mixture of two coherent states as one observes in Fig. \ref{fig2s}\hyperref[fig2s]{(b)}. 

\section{PHASE-LOCKED SQUEEZED LASING VIA EXTERNAL DRIVING}
\label{sm3}
\begin{figure}[t]
\centering
\includegraphics[width=0.65\linewidth]{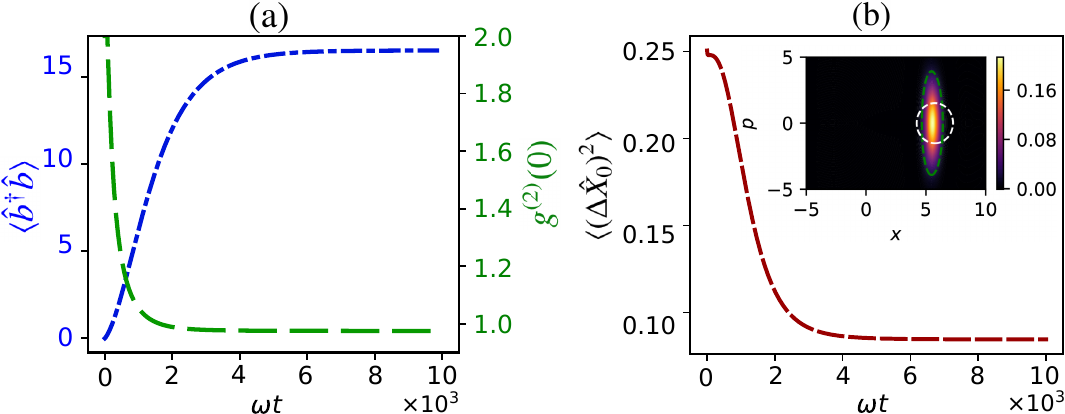}
\caption{Numerical solution of ME in Eq. \eqref{MEF}: $(a)$ Time evolution of excitations and second-order correlation function, $g^{(2)}(0)$. $(b)$ Time evolution of quadrature fluctuation $\langle (\Delta \hat{X}_{\theta})^2 \rangle$, for $\theta=0$. Inset plot: Wigner function $W(x,p)$ at $t\rightarrow\infty$. The green dashed lines indicate the contour at $0.1\max[W(x,p)]$, while the white dashed lines correspond to the same contour for a coherent field with identical coherent amplitude. Parameters: $\lambda_1=0.04\omega$, $J=0.3\omega$, $\Omega_d=5.5\times10^{-3}\omega$, $\phi=\pi/2$, $\chi=1$, $\Gamma_{j}=10^{-1}\omega$, $\gamma_m=1.5\times10^{-4}\omega$, $\gamma'_m=12\gamma_m$, $\bar{n}_{j}^{s}=10^{-5}$, $\bar{n}_{m}=10^{-3}$.}
\label{fig6}
\end{figure}

In a conventional laser, the system may select a phase that remains stable over extremely long times. For instance, one can introduce an external perturbation, such as a weak coherent field, which explicitly breaks the phase symmetry. To illustrate this effect, Fig. \ref{fig6} summarizes the impact of adding a small coherent drive applied to the primary spins, with a well-defined phase, described by the Hamiltonian term $\hat{H}_d = \Omega_d \left( \hat{B} e^{-i\phi} + \hat{B}^{\dagger} e^{i\phi} \right)$. Here, $\Omega_d$ and $\phi$ denote the amplitude and phase of the external field, respectively, and the operator $\hat{B}$ is the same as the one appearing in $\hat{H}_{\mathrm{eff}}^{A}\equiv\hat{H}_2(\nu/\omega=2)$ (defined in Eq. \eqref{eq:heff1_l}).

By including the external drive, the dissipative dynamics of the hybrid quantum system can be described by an effective ME for the full system defined in Section \ref{sec_hBNa}:
\begin{eqnarray}\label{MEF}
     \frac{d\hat{\rho}}{dt}&=&-i[\hat{H}_{\text{eff}}^A+\hat{H}_d,\hat{\rho}]+\gamma'_m\mathcal{L}_{\hat{B}}[\hat{\rho}]+\gamma_m\left(1+\bar{n}_{m}\right)\mathcal{L}_{\hat{b}}[\hat{\rho}]+\gamma_m\bar{n}_{m}\mathcal{L}_{\hat{b}^\dagger}[\hat{\rho}]+\sum_{j=1}^2\Gamma_{j}\left(1+\bar{n}_{j}^{s}\right)\mathcal{L}_{\hat{\sigma}_{j}^{-}}[\hat{\rho}]+\Gamma_{j}\bar{n}_{j}^{s}\mathcal{L}_{\hat{\sigma}_{j}^{+}}[\hat{\rho}].
\end{eqnarray}
Here, the damping of the squeezed mode $\hat{B}$ is described by the decay rate $\gamma'_{m}$, whose derivation is presented in Sec. \ref{sec4}.

Analogous to the results in Sec. \ref{sec_phase_control}, Fig. \ref{fig6}\hyperref[fig6]{(a)} shows the buildup of the phonon population together with the convergence of $g^{(2)}(0)$ towards unity, indicating the emergence of coherent phonon lasing from an initially thermal state. Small residual deviations from $g^{(2)}(0)=1$ are associated with the squeezing character of the phonon field.  

In Fig. \ref{fig6}\hyperref[fig6]{(b)}, the quadrature variance $\langle (\Delta \hat{X}_{0})^2 \rangle$ evolves into the squeezed regime and stabilizes below the vacuum limit, evidencing steady-state squeezing. Correspondingly, the Wigner function displays the characteristic elongated profile of a squeezed coherent state, confirming the realization of squeezed phonon lasing.

\section{QUANTUM-ENHANCED SENSING WITH SQUEEZED PHONON LASER}\label{appendix_metrology}

The emergence of the squeezed-phonon-lasing regime not only gives rise to nonclassical phonon emission but also may open new opportunities for quantum-enhanced metrology. Since the local driving amplitude, $\varepsilon\equiv\chi \nu/2$, acts as an effective squeezing parameter (see details in Appendix \ref{sm4}), small variations of $\chi$ produce significant modifications in the steady-state phonon statistics and phase-space distribution. Consequently, the MO becomes highly sensitive to the local driving strength, suggesting that the squeezed-lasing regime may serve as a resource for precision sensing.

To quantify a signature of enhanced sensitivity, we resort in quantum estimation theory~\cite{paris2009quantum}. The cornerstone of quantum estimation theory is the quantum Fisher information (QFI)~\cite{montenegro2025review, giovannetti2004quantum, giovannetti2006quantum, giovannetti2011advances, paris2009quantum} $\mathcal{F}_{Q}$. From a geometric perspective, the QFI induces the Bures Riemannian metric on the manifold of quantum states and therefore quantifies the statistical distinguishability between neighboring quantum states that differ by an infinitesimal variation of an encoded parameter~\cite{braunstein1994statistical}. Equivalently, the QFI determines the infinitesimal statistical distance between the states, say $\hat{\rho}(\chi)$ and $\hat{\rho}(\chi+d\chi)$ through $ds^{2}=\frac{1}{4}\mathcal{F}_{Q}(\theta)d\chi^{2}$~\cite{sidhu2020geometric}. Consequently, a larger value of $\mathcal{F}_{Q}$ implies that an infinitesimal variation of the parameter produces a greater statistical separation between the corresponding quantum states, leading to a higher achievable estimation precision. This connection is made quantitative by the quantum Cram\'{e}-Rao bound~\cite{helstrom1967minimum, helstrom1976quantum, holevo2011probabilistic, vantrees2004detection, Cramer},
\begin{equation}
\mathrm{Var}[\chi]\ge\frac{1}{M\mathcal{F}_Q(\chi)},
\label{eq:QCRB}
\end{equation}
where $M$ denotes the number of independent measurements, $\mathrm{Var}[\chi]$ is the variance of the estimator of $\chi$ and $\mathcal{F}_Q(\chi)$ is the QFI with respect to unknown parameter $\chi$. For the mixed quantum states, the QFI can be written as~\cite{paris2009quantum},
\begin{equation}
\mathcal{F}_Q=2\sum_{n,m}\frac{\left|\langle\psi_m|\partial \rho_\chi/\partial\chi|\psi_n\rangle
\right|^2}{\epsilon_m+\epsilon_n},\qquad
\epsilon_m+\epsilon_n\neq0,
\label{eq:QFI_spectral}
\end{equation}
where $\rho_\chi=\sum_n\epsilon_n|\psi_n\rangle\langle\psi_n|$ with $\epsilon_n$ and $\ket{\psi_n}$ denote the $n$th eigenvalue and eigenvector of $\rho_\chi$, respectively. Here, $\rho_\chi$ is the joint spins-mechanical steady-state assuming $\chi$ to be unknown, whereas all the rest of quantities are known. 

\begin{figure}[t]
\centering
\includegraphics[width=0.75\linewidth]{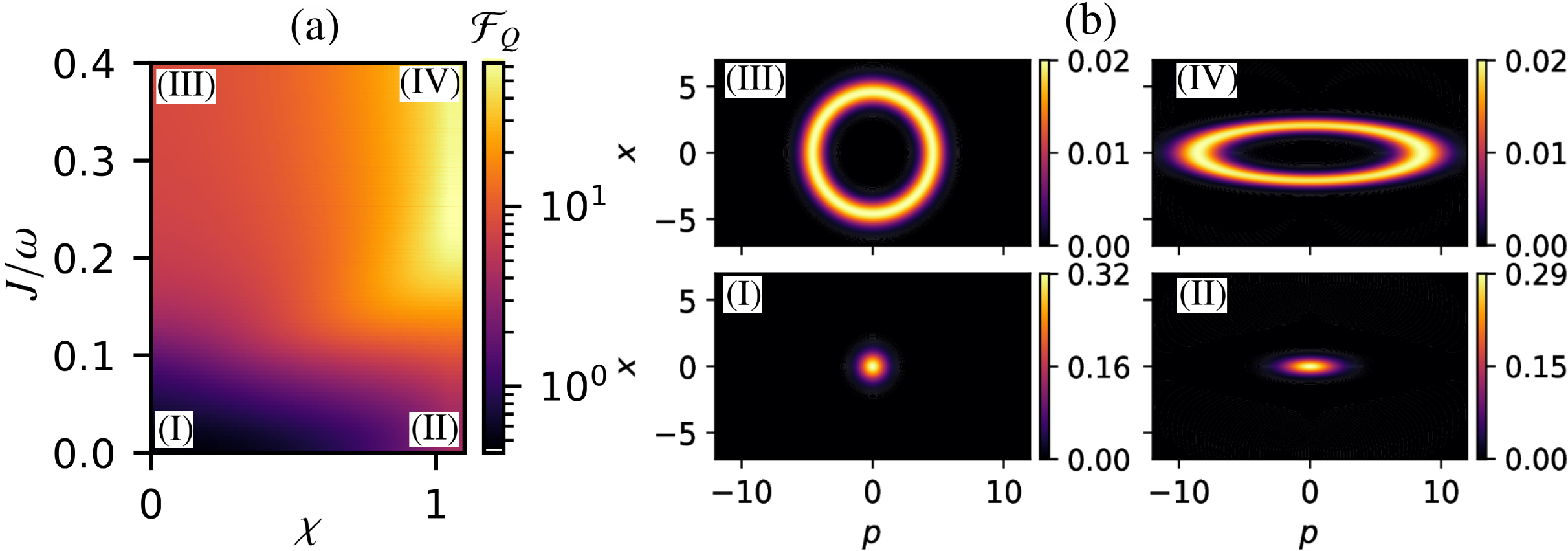}
\caption{$(a)$ Quantum Fisher information $\mathcal{F}_Q$ with respect to $\chi$ as a function of $J$ and $\chi$ for $\nu=2\omega$. $(b)$ Wigner functions $W(x,p)$ of the MO corresponding to the cases shown in panel (a): (I) $J/\omega=0$, $\chi=0$; (II) $J/\omega=0$, $\chi=1$; (III) $J/\omega=0.4$, $\chi=0$; and (IV) $J/\omega=0.4$, $\chi=1$. Other parameters are the same as in Fig.~\ref{fig3}.}
\label{figq}
\end{figure}

In Fig.~\ref{figq}\hyperref[figq]{(a)}, we present the steady-state QFI of the joint spins-mechanical state with respect to the squeezing parameter $\chi$ as a function of both $\chi$ and the spin-spin drive amplitude $J$. Note that, because $\chi = 2\varepsilon/\nu$ with $\nu = 2\omega$, determining the sensing precision limits of $\chi$ is equivalent to determining the precision limits for $\varepsilon$, the AC drive amplitude in our scheme. As shown in Fig.~\ref{figq}\hyperref[figq]{(a)}, for a fixed value of $\chi$, the QFI increases with increasing $J$ before reaching a plateau. Conversely, for a fixed value of $J$, the QFI rises sharply as $\chi$ increases. This behavior is particularly pronounced around $J \approx 0.2\omega$, where the strongest changes in $g^{(2)}(0)$ occur. This parameter regime coincides with the onset of stable steady-state squeezed lasing, see Fig.~\ref{fig3}\hyperref[fig3]{(b)} in the main text. As the system evolves from the conventional phonon-lasing regime ($\chi = 0$) to the squeezed-phonon-lasing regime ($\chi > 0$), the QFI increases by more than an order of magnitude, revealing a substantial enhancement in the sensitivity of the steady state to infinitesimal variations in $\chi$.

In Fig.~\ref{figq}\hyperref[figq]{(b)}, we show four representative steady-state Wigner functions corresponding to the points marked in Fig.~\ref{figq}\hyperref[figq]{(a)}. As seen from Fig.~\ref{figq}\hyperref[figq]{(b)}, the squeezed-phonon-lasing regime exhibits the largest QFI values. This result demonstrates that the same physical mechanism responsible for generating nonclassical phonon emission also enhances the statistical distinguishability of neighboring quantum states, thereby providing a significant metrological advantage.

\twocolumngrid

\bibliography{biblio2}

\end{document}